\documentclass[iop]{emulateapj}

\usepackage{color}

\newcommand{\ltsimeq}{\raisebox{-0.6ex}{$\,\stackrel
        {\raisebox{-.2ex}{$\textstyle <$}}{\sim}\,$}}

\newcommand{\gtsimeq}{\raisebox{-0.6ex}{$\,\stackrel
        {\raisebox{-.2ex}{$\textstyle >$}}{\sim}\,$}}

\shortauthors{Helton et al.}

\shorttitle{``Late Time Observations of Classical Novae''}

\submitted{Submitted to ApJ - May 8, 2012}

\revised{June 14, 2012}
\accepted{June 17, 2012}

\begin{document}

\title{Elemental Abundances in the Ejecta of Old Classical Novae from Late-Epoch \textit{Spitzer} Spectra} 

\author{
L.\ Andrew Helton\altaffilmark{1}, 
Robert D.\ Gehrz\altaffilmark{2},
Charles E.\ Woodward\altaffilmark{2}, 
R. Mark Wagner\altaffilmark{3},
William D.\ Vacca\altaffilmark{1},
Aneurin Evans\altaffilmark{4},
Joachim Krautter\altaffilmark{5},
Greg J.\ Schwarz\altaffilmark{6}, 
Dinesh P.\ Shenoy\altaffilmark{2},
Sumner Starrfield\altaffilmark{7}
} 

\altaffiltext{1}{SOFIA Science Center, USRA, NASA Ames Research Center, M.S. N232-11, Moffett Field, CA 94035, USA, \textit{ahelton@sofia.usra.edu}}

\altaffiltext{2}{Minnesota Institute for Astrophysics, School of Physics and Astronomy, 116 Church Street S.E., University of Minnesota, Minneapolis, Minnesota 55455, USA}

\altaffiltext{3}{Department of Astronomy, The Ohio State University, 140 W. 18$^{\mathrm{th}}$ Ave., Columbus, OH, 43210, USA}

\altaffiltext{4}{Astrophysics Group, Keele University, Keele, Staffordshire ST5, 5BG, UK} 

\altaffiltext{5}{Landessternwarte - Zentrum f\"ur Astronomie der Universit\"at, K\"onigstuhl, D-69117, Heidelberg, Germany} 

\altaffiltext{6}{American Astronomical Society, 2000 Florida Ave, NW, Suite 400, Washington, DC 20009, USA}

\altaffiltext{7}{School of Earth and Space Exploration, Arizona State University, P.O. Box 871404, Tempe, AZ 85287, USA}




\begin{abstract}

We present \textit{Spitzer Space Telescope} mid-infrared IRS spectra, supplemented by ground-based optical observations, of the classical novae V1974 Cyg, V382 Vel, and V1494 Aql more than 11, 8, and 4 years after outburst respectively. The spectra are dominated by forbidden emission from neon and oxygen, though in some cases, there are weak signatures of magnesium, sulfur, and argon. We investigate the geometry and distribution of the late time ejecta by examination of the emission line profiles. Using nebular analysis in the low density regime, we estimate lower limits on the abundances in these novae. In V1974 Cyg and V382 Vel, our observations confirm the abundance estimates presented by other authors and support the claims that these eruptions occurred on ONe white dwarfs. We report the first detection of neon emission in V1494 Aql and show that the system most likely contains a CO white dwarf. 

\end{abstract} 


\clearpage


\section{Introduction}
\label{sec:intro}

A classical nova (CN) explosion results from a thermonuclear runaway (TNR) on the surface of a white dwarf (WD) that has accreted matter from a dwarf secondary star in a close binary system.  The elemental composition of the ejected material depends both on the degree of thermonuclear processing during the TNR and the amount and composition of the material dredged up from the underlying WD. CO-type nova eruptions are thought to arise on the surface of relatively low-mass (M$_{\mathrm{WD}} < 1.2$ M$_{\odot}$) WDs composed primarily of carbon and oxygen, whereas evidence suggests that ONe-type (or simply ``neon'') novae arise from TNRs on high-mass (M$_{\mathrm{WD}} > 1.2$ M$_{\odot}$) oxygen, neon, and magnesium-rich WDs \citep{Gehrz08b}. 

Infrared (IR) observations are critical for accurately constraining photoionization models of CN ejecta \citep[e.g.][]{Helton10a} since diagnostics derived from only the optical or UV fail to sample a broad enough range of ionization and excitation levels in the emission lines. IR observations may also allow the determination of abundances in systems that are subject to such a high degree of extinction that few emission features are detected in the UV or optical \citep[e.g., V1721 Aql;][]{Hounsell11}. In addition, late-epoch IR observations of CNe are particularly valuable for revealing metals in the ejecta through emission lines with very low critical densities. In some cases, these may arise from elemental species with few or weak transitions at ultraviolet or optical wavelengths.

Ideally, the abundances of metals are determined by comparing the flux of the forbidden metal lines to hydrogen recombination lines present at the same time. However, at late epochs ($\gtsimeq 1000$ days post-outburst), the infrared spectra are devoid of hydrogen recombination line emission, and thus, we are limited to estimates of relative abundances, unless a proxy for the hydrogen number density, $n_{H}$, is available. In these cases one may use early post-eruption measurements to constrain the total mass of the hydrogen expelled by the nova allowing the calculation of the metal abundances relative to hydrogen.  

In this paper, we report on mid-infrared observations CNe in their late nebular stages ($>4$ years post-outburst), including V1974 Cyg, V382 Vel, and V1494 Aql. Table \ref{tab:properties} lists the targets with some of their basic observational characteristics. In \S\ref{sec:OldNovae_Obs}, we detail the \textit{Spitzer} and supporting optical observations. This is followed by an overview of the abundance determination methods and parameter selection in \S\ref{sec:Abundances}. Section \ref{sec:Analysis} then presents a detailed analysis of each of the targets, which includes examination of the emission lines to characterize the nebular environment as well as abundance calculations. In \S\ref{sec:Discussion}, we place the derived abundances in context with results for other systems of both the CO and ONe classes. Finally, we present our conclusions in \S\ref{sec:Conclusions}.


\begin{deluxetable*}{lccccccccc}
\tabletypesize{\scriptsize}
\setlength{\tabcolsep}{0.05in}

\tablewidth{0pt}
\tablecaption{Fundamental Properties \label{tab:properties}}

\tablehead{
\colhead{Target}& \colhead{RA (J2000.0)}& \colhead{Dec (J2000.0)}& \colhead{t$_{max}$ (UT)}&
\colhead{m$_{V,max}$}& \colhead{t$_{2}$\tablenotemark{a}}& \colhead{t$_{3}$\tablenotemark{b}}& \colhead{Speed Class}&
\colhead{Type\tablenotemark{c}}& \colhead{Dust?}
}

\startdata

V1500 Cyg& 21$^{h}$11$^{m}$36\fs61& +48\degr09'01\farcs9& 1975 Aug. 29.5& 1.8& 2.9& 3.6& Very Fast& Hybrid\tablenotemark{d}; ONe?& No\\
NQ Vul& 19$^{h}$29$^{m}$14\fs68& +20\degr27'59\farcs7& 1976 Oct. 21.8& 6.5& 23& 53& Fast&\ion{Fe}{2}; ONe?& Yes\\
V1668 Cyg& 21$^{h}$42$^{m}$35\fs31& +44\degr01'55\farcs0& 1978 Sep. 12.2& 6.0& 12.2& 24.3& Fast& \ion{Fe}{2}; CO& Yes\\
V1974 Cyg& 20$^{h}$30$^{m}$31\fs66& +52\degr37'51\farcs3& 1992 Feb. 20.8& 4.2& 16-24& 47& Fast& \ion{Fe}{2}; ONe& No\\
V382 Vel& 10$^{h}$44$^{m}$48\fs37& -52\degr25'30\farcs6& 1999 May 22.4& 2.3& 4.5-6& 9-12.5& Very Fast& \ion{Fe}{2}; ONe& No\\
V1494 Aql& 19$^{h}$23$^{m}$05\fs28& +04\degr57'21\farcs6& 1999 Dec. 03.4& 4.0& 6.6& 16& Very Fast& \ion{Fe}{2}; CO?& No\\[-2.5ex]

\enddata
\tablenotetext{a}{t$_{2}$ is the time (in days) it takes for the nova to decline 2 visual magnitudes from its maximum.}
\tablenotetext{b}{t$_{3}$ is the time (in days) it takes for the nova to decline 3 visual magnitudes from its maximum.}
\tablenotetext{c}{The first entry is for the classification of the spectrum as presented by \citet{Williams92}. The second entry identifies the system based on the underlying WD composition.}
\tablenotetext{d}{``Hybrid'' novae exhibit spectra that transition from \ion{Fe}{2}-type to He/N-type.}

\end{deluxetable*}



\section{Observations and Reduction}
\label{sec:OldNovae_Obs}

\subsection{\textit{Spitzer} Mid-Infrared Spectroscopy}
\label{sec:OldNovae_MIR}

We observed nine old CNe (QU Vul, GK Per, V1500 Cyg, NQ Vul, V1668 Cyg, V705 Cas, V1974 Cyg, V382 Vel, and V1494 Aql) using the IRS instrument \citep{Houck04} on board the \textit{Spitzer} Space Telescope \citep[\textit{Spitzer};][]{Werner04,Gehrz07} as part of our \textit{Spitzer} CN monitoring program (program identification numbers (PIDs) 122, 124 (P.I. RDG), 30076, and 40060 (P.I. AE)). The novae were selected for study to assess the late stage evolution of optically bright (m$_{\mathrm{V}} \ltsimeq 5$~mag) outbursts as the systems returned to quiescence. The observations of V1500 Cyg, NQ Vul, and V1668 Cyg resulted in null detections and will be discussed briefly below. Our analysis of QU Vul was presented by \citet{Gehrz07}. Analysis of the data on both the dusty nova V705 Cas and the very old nova GK Per will be presented elsewhere. A summary of the \textit{Spitzer} observations presented in the present work is provided in Table \ref{tab:sp_obs}. 

The observations were conducted using the short wavelength (5 -- 15~\micron)  low resolution module (SL), the  short wavelength (10 -- 19~\micron)  high resolution module (SH), the long  wavelength (19 -- 37~\micron) high resolution module (LH), and, in some cases, the long wavelength (14 -- 38~\micron) low resolution module (LL).   The resolving power for SL and LL is $R$ = $\lambda/\Delta\lambda \sim 60-120$ and R $\sim$ 600 for SH and LH. Observations  in PIDs 122 and 124 utilized visual Pointing Control Reference Sensor \citep[PCRS,][]{Mainzer98} peak-up on nearby isolated stars to ensure proper placement of the target in the narrow IRS slits, while PIDs 30076 and 40060 used on-source peak-up. The time on source varied by module and program and is provided in Table \ref{tab:sp_obs}. 

IRS Basic Calibrated Data (BCD) products were produced by the \textit{Spitzer} Science Center (SSC) IRS pipelines version v15.3.0, v16.1.0, and v17.2.0 depending upon the module and epoch of observation (see Table \ref{tab:sp_obs}). Details of the calibration and raw data processing are specified in the IRS Pipeline Description Document, v1.0.\footnote{http://ssc.spitzer.caltech.edu/irs/dh/PDD.pdf}.  Bad pixels were interpolated over in individual BCDs using bad pixel masks provided by the SSC.  For all PIDs, multiple data collection events were obtained at two different positions on the slit using \textit{Spitzer's} nod functionality. For PIDs 122 and 124, sky subtraction was possible only for the low-res observations as no dedicated sky observations were performed for the SH and the LH mode observations. For the low-res modules, the median combined two-dimensional spectra were differenced to remove the background flux contribution. PIDs 30076 and 40060 specified separate offset observations of ``blank'' sky from which background images were derived by median combining the two-dimensional spectral images. These were then subtracted from the  on-source BCDs. Spectra were then extracted from the two-dimensional images with the \textit{Spitzer} IRS Custom Extraction Software (SPICE\footnote{http://ssc.spitzer.caltech.edu/dataanalysistools/tools/spice/}; version 1.4.1). Based on the peak expansion velocities, distances, and ages at the time of the \textit{Spitzer} observations (see \S\ref{sec:Analysis} below), we expect the sources to be unresolved with the IRS. Hence, we conducted the extraction using the default point source extraction widths. The extracted spectra were combined using a weighted linear mean into a single output data file. Due to the low continuum signal, it was not necessary to de-fringe the data. Errors were estimated from the standard deviation of the fluxes at each wavelength bin. In the case of the high resolution data, this results in an overestimation of the errors due to variable slit losses between nod positions, which were typically higher than the errors expected based on examination of the continuum. Fitting of the spectral lines was performed using a non-linear least squares Gaussian  routine \citep[the Marquardt method;][]{Bev92} that fits the line  center, line amplitude, line width, continuum level, and continuum slope. All of the lines were resolved in the high resolution modules. Those lines exhibiting castellated (saddle-shaped) structure were fit with multiple Gaussians in order to accurately assess the line widths and fluxes.


\begin{deluxetable*}{lcccccccc}
\tabletypesize{\scriptsize}
\setlength{\tabcolsep}{0.05in}

\tablecaption{\textit{Spitzer} Observations \label{tab:sp_obs}}

\tablehead{
\colhead{}& \colhead{}& \colhead{}& \colhead{}& \colhead{}& \colhead{Observation}& \colhead{JD}& \colhead{$\Delta$t\tablenotemark{a}}& \colhead{On-Source}\\
\colhead{Target}& \colhead{PID}& \colhead{Modules}& \colhead{AOR}& \colhead{Pipeline}& \colhead{Date (UT)}& \colhead{(-2450000)}& \colhead{(Days)}& \colhead{Time\tablenotemark{b} (sec)}
}

\startdata

V1974 Cyg& 124& SL, SH, LH& 5043968& 15.3.0& 2003 Dec 17.4& 2990.9& 4318& 24/84/84\\
& 30076& SL, SH, LH& 17732096/7216\tablenotemark{c}\tablenotemark{d}& 15.3.0/17.2.0& 2006 Oct 18.4& 4026.9& 5354& 72/72/168\\
& 40060& SL, SH, LH& 22267648/7904& 16.1.0/17.2.0& 2007 Aug 30.2& 4342.7& 5669& 140/300/600\\
V382 Vel& 122& SL, SH, LH& 5021696& 15.3.0& 2004 May 13.6& 3139.1& 1820& 60/72/72\\
& 30076& SL, SH, LH& 17733120/8240\tablenotemark{d}& 16.1.0/15.3.0/17.2.0& 2007 Mar 26.2& 4185.7& 2867& 600/72/72\\
& 40060& All& 22269440/9696& 17.2.0& 2008 Feb 25.4& 4521.9& 3203& 140/300/140/600\\
V1494 Aql& 122& SL, SH, LH& 5022464& 15.3.0& 2004 Apr 14.3& 3109.8& 1594& 60/72/72\\
& 30076& SL, SH, LH& 17732864/7984\tablenotemark{d}& 16.1.0/17.2.0& 2007 Jun 08.2& 4259.7& 2744& 48\tablenotemark{e}/96/224\\
& 40060& All& 22268928/9184& 16.1.0/17.2.0& 2007 Oct 12.3& 4385.8& 2870& 140/60/140/140\\
V1500 Cyg& 124& SL, SH, LH& 5047040& 15.3.0& 2003 Dec 16.8& 2990.3& 10335& 60/72/72\\
NQ Vul& 124& SL, SH, LH& 5040384& 15.3.0& 2004 Apr 16.6& 3112.1& 10039& 60/72/72\\
V1668 Cyg& 124& SL, SH, LH& 5047552& 15.3.0& 2003 Dec 15.5& 2989.0& 9227& 60/72/72\\[-2.5ex]

\enddata

\tablenotetext{a}{Time since maximum.}
\tablenotetext{b}{On-source integration times are given in SL/SH/LL/LH order (see \S\ref{sec:OldNovae_MIR}).}
\tablenotetext{c}{The second number is the last four digits of the background AOR.}
\tablenotetext{d}{For these observations a background exposure was acquired for the LH module only.}
\tablenotetext{e}{For this observation, the on-source integration time in the SL module was 48 seconds in the 5.2 -- 8.7 \micron\ range and 72 sec in the 7.4 -- 14.5 \micron\ range.}
\end{deluxetable*}


\subsection{Optical Spectroscopy}
\label{sec:OldNovae_Optical}

Both V1974 Cyg and V1494 Aql were observed with the Boller \& Chivens Spectrograph\footnote{http://james.as.arizona.edu/\~{}psmith/90inch/90finst.html} (B\&C) at the Steward Observatory 2.3~m Bok telescope on Kitt Peak, Arizona as part of our on-going CN optical monitoring campaign. Details of the observations are presented in Table \ref{tab:opt_obs}. A 400 line/mm first order grating was employed and when combined with a 1\farcs5 wide entrance slit yielded a spectral resolution at the detector about $\sim$6 \AA.  Two grating tilts were required to cover the entire spectral region between 3550-9375 \AA.  Order separation filters were utilized to avoid contamination of the spectra by adjacent orders.  

In addition, V1494 Aql was observed at two early epochs (Days 144 and 185) with the Boller and Chivens CCD Spectrograph (CCDS) attached to the MDM Observatory 2.4~m Hiltner telescope on Kitt Peak, Arizona.  These observations are also summarized in Table 3.  A 150 lines/mm grating in first order was utilized and when combined with a 1\farcs0 wide entrance slit yielded a spectral resolution of $\sim$8 \AA.  The spectra cover the region 3200--6800 \AA\ at a dispersion of 3 \AA/pixel.  No order separation filter was used.  


\begin{deluxetable*}{lccccccc}
\tabletypesize{\scriptsize}
\setlength{\tabcolsep}{0.05in}

\tablewidth{0pt}
\tablecaption{Optical Observations\label{tab:opt_obs}}

\tablehead{
\colhead{}& \colhead{}& \colhead{$\lambda_{\mathrm{cent}}$}& \colhead{}& \colhead{Integration}& \colhead{Observation}& \colhead{JD}& \colhead{$\Delta$t\tablenotemark{a}}\\
\colhead{Target}& \colhead{Facility}& \colhead{(\AA)}& \colhead{Filter}& \colhead{Time (s)}& \colhead{Date (UT)}& \colhead{(-2450000)}& \colhead{(Days)}
}

\startdata

V1974 Cyg& Bok 2.3-m& 5200& \nodata& 5100& 2006 Sep 28.3& 4006.8& 5334\\
& Bok 2.3-m& 5200& UV36& 2700& 2006 Sep 30.2& 4008.7& 5335\\
& Bok 2.3-m& 5200& UV36& 2700& 2007 Oct 20.1& 4394.6& 5720\\
& Bok 2.3-m& 5200& UV36& 300& 2008 Sep 22.3& 4731.8& 6059\\
& Bok 2.3-m& 7675& Y48& 300& 2008 Sep 23.3& 4732.8& 6060\\
& Bok 2.3-m& 7675& Y48& 1500& 2008 Sep 24.3& 4733.8& 6061\\[0.8ex]
\hline\\[-1.4ex]
V1494 Aql& Hiltner 2.4-m& 5000& \nodata& 90& 2000 Apr 25.5& 1659.0& 143\\
& Hiltner 2.4-m& 5000& \nodata& 120& 2000 Jun 4.5& 1701.0& 185\\
& Bok 2.3-m& 4770& \nodata& 1800& 2004  Jun 25.3& 3181.8& 1666\\
& Bok 2.3-m& 7830& Y48& 1800& 2004  Jun 25.3& 3181.8& 1666\\[-2.5ex]

\enddata

\tablenotetext{a}{V1974 Cyg - Calculated from t$_{0} = $ JD 2448673.3; V1494 Aql - Calculated from t$_{0} = $ JD 2451515.9}

\end{deluxetable*}


The data were reduced in IRAF\footnote{IRAF is distributed by the National Optical Astronomy Observatories, operated by the Association of Universities for Research in Astronomy, Inc., under cooperative agreement with the National Science Foundation.} using standard optical spectral data reduction procedures. For each set of observations, three (CCDS) and five or more (B\&C) spectra were obtained and combined with median filtering to allow the removal of cosmic rays. Wavelength calibration was performed using either HeArNe (B\&C) or HgArNe (CCDS) calibration lamps.  Corrections for pixel--to--pixel variations in response (flatfield) was accomplished using a high intensity quartz--halogen lamp in each instrument. In the case of the B\&C spectra, fringing occurred beyond 7000 \AA. These fringes can be problematic when one is trying to interpret flat-fielded spectra of bright targets, but in the case of V1974 Cyg, the fringes were weaker than the relatively high noise level and did not have an appreciable effect. Spectra of the spectrophotometric standard stars were obtained at an airmass similar to that of the target novae to correct for instrumental response and provide flux calibration. Line fluxes were measured from the reduced spectra using the same method as for the \textit{Spitzer} spectra.

\section{Abundance Determination}
\label{sec:Abundances}

There are two general methods for determining ejecta abundances in novae -- nebular analysis and photoionization modeling. The choice of technique depends upon the properties of the expanding material within the line emitting region and the available data. 

Mid-IR forbidden emission lines in novae are produced through spontaneous emission from excited states populated through collisions within the ejecta. For very low density gas, the spontaneous emission of a photon from an excited atom will occur before collisional de-excitation. In this low density limit, we assume that every collisional excitation results in subsequent photon emission such that the rate of emission is directly related to the rate of collisions within the gas. In this regime, standard nebular analysis techniques (``detailed balancing'') such as those described by \citet{Osterbrock06} are valid and allow determination of the abundances within the line emitting region. The uncertainty in the abundance solution obtained through detailed balancing arguments is contingent upon the accuracy of the electron temperature and density. Further, the derived abundances for each element are necessarily lower limits since there may exist additional emission lines from unseen ionization states that fall outside the bandpass of the observed spectral region and are not accounted for in these calculations. This method makes no attempt to ``correct'' for the missing populations producing these unobserved lines. 

The density at which significant fractions of the excited ions no longer have time to emit a photon before they suffer de-excitation through particle collisions is called the critical density, \textit{n}$_{crit}$. If the electron density, $n_{e}$, is lower than n$_{crit}$ so that collisional de-excitation is negligible, then it is straightforward to determine the population of the upper excitation state from the observed line intensities. At higher densities, the rate of emission is no longer directly related to the rate of excitation since the lines are subjected to collisional damping. The simplistic assumptions made in the low density limit are no longer valid and one must use the full level balancing equations to properly estimate the population of the emitting species. If all ionization states actually present in the ejecta are not observed, then it may be necessary to use photoionization codes \citep[e.g., Cloudy,][]{Fer98} to facilitate an accurate accounting of the abundances.

Photoionization analysis cannot be used to reliably estimate the abundances in the ejecta at late times in CNe since this method assumes an illuminating source that provides a steady flux of incident photons leading to radiative excitation. In the case of each of the CNe in our sample, the central source had ``turned off'' at least 1000 days prior to our \textit{Spitzer} observations \citep{Schwarz11}, which means that the photon flux from the central system is low, of order a few L$_{\odot}$, and no longer produces a substantial flux of UV photons capable of significant photoionization and radiative excitation. Hence, these photoionization models would not yield physically meaningful results.

The ejecta in the CNe presented here have all undergone many years of expansion at velocities $\gtsimeq 10^{3}$~km s$^{-1}$. Based on our estimates for the shell geometries, masses of ejected hydrogen, and electron temperatures, we determined if the ejecta densities had declined below the critical densities for the observed transitions. If so, we used standard nebular analysis techniques following the methodology of \citet{Osterbrock06} as demonstrated in our earlier analysis of QU Vul \citep{Gehrz08a}. 

We derived the abundances of metals in the ejecta by number relative to hydrogen by comparison of the sum of the populations of an observed species to the total number of hydrogen atoms. We then compared these ionic abundances to the total solar abundances compiled by \citet{Asplund09}, where the logarithmic solar abundances, log(N$_{\mathrm{X}}$/N$_{\mathrm{H}}$)$_{\odot}$, are: He = -1.07; O = -3.31; Ne = -4.06; Mg = -4.40; S = -4.88; Ar = -5.60; and Fe = -4.50.

A limitation of this approach is that the spectra presented here no longer contain any signatures of hydrogen recombination emission from the ejecta. This absence means that we are unable to make a direct comparison between the populations implied by the observed emission lines to the amount of hydrogen within the ejecta. Thus, we are forced to rely on estimations of the ejected hydrogen mass made at earlier epochs. This source of uncertainty dominates our estimates of the elemental abundances with respect to hydrogen. The accuracy of this method of determining abundances presumes that the amount of hydrogen in the ejected shell has remained roughly constant since the early stages of development of the ejecta following outburst and was not significantly enhanced, for example, by a massive persistent  post-outburst wind. 

\subsection{Parameter Selection}
\label{sec:Params}

For a fully ionized medium, \textit{n}$_{e}$ is assumed to be approximately equal to the hydrogen density of the gas, \textit{n}$_{H}$. In a strict sense, this is the lower limit to \textit{n}$_{e}$ as there is also a contribution to the electron population by the ionized metals. In the case of ejecta with a high metal abundance (e.g., $>100$ times solar), this contribution to the electron population may be significant, resulting in a higher value for \textit{n}$_{e}$ than we assume here. We neglect this contribution to \textit{n}$_{e}$ as its impact on the derived abundances is small compared to error introduced due to the assumed ejection mass (as described below).

A direct estimate of the electron temperature in the ejecta, $T_{e}$, can be obtained using the ratio of [\ion{O}{3}] $\lambda$4363 \AA\ to [\ion{O}{3}] $\lambda\lambda$4959, 5007 \AA\ \citep{Osterbrock06}. As the temperatures in the nebula decline ($\ltsimeq 10^{4}$ K), the strength of the 4363 \AA\ transition decreases rapidly, making the use of the [\ion{O}{3}] line ratios for temperature diagnostics difficult. 

In contrast, IR fine structure lines, which arise through collisional excitation in low density gas, can easily be excited at low temperatures due to their much lower excitation potentials. Combinations of IR and optical lines, e.g., [\ion{Ne}{5}] at 24.3 \micron\ and 3426 \AA, can provide accurate temperature estimates if the reddening is known. Otherwise one may use far-IR line ratios such as [\ion{O}{1}] 145 \micron\ relative to 63 \micron\ \citep[][and references therein]{Dinerstein95}. In the present data, the optical and far-IR lines could not be used. So instead, we derived estimates for the temperature based upon the ratio of [\ion{Ne}{5}] 14.32 to 24.30 \micron, when available.

In the event that none of the above methods for estimating $T_{e}$ were available, we derived estimates by comparison with other CNe taking into consideration the relative ages of the systems.Cooling by IR fine structure lines could be very efficient provided that the densities in the ejecta were lower than the critical densities of the prominent IR transitions \citep{Ferland84}, and particularly if the ejecta are metal-rich \citep{Smits91}. Support for low nebular temperatures at late times was demonstrated for QU Vul by \citet{Gehrz08a}, who estimated the temperature of the ejecta to be $\simeq 1000$ K at 7.6 years post outburst based on narrow-band imaging obtained by \citet{Krautter02}. So, for those systems that have no other reliable temperature estimate, we calculate the abundances using electron temperatures of 1000 K and 5000 K, which we consider to be reasonable extremes based on their ages and the likelihood that the ejecta have undergone substantial cooling by the IR fine structure lines.

The effective collision strength, $\Upsilon_{ij}$, is dependent on the assumed $T_{e}$ of the ejecta. The adopted values of $\Upsilon_{ij}$, spontaneous emission coefficients (A$_{ij}$), and the sources from which they were obtained are given in Table \ref{tab:lineparams}.


\begin{deluxetable*}{lccccccccccc}
\tabletypesize{\scriptsize}
\setlength{\tabcolsep}{0.05in}

\tablewidth{0pt}
\tablecaption{Line Parameters \label{tab:lineparams}}

\tablehead{
\colhead{Ion}& \colhead{$\lambda$}& \colhead{Term}& \colhead{A$_{ij}$}& \multicolumn{3}{c}{$\Upsilon_{ij}$}& \colhead{}& \multicolumn{3}{c}{$n_{crit}$ (cm$^{-3}$)}& \colhead{References\tablenotemark{a}\tablenotemark{b}}\\[0.5ex]
\cline{5-7} \cline{9-11}\\[-1.8ex]
\colhead{}& \colhead{(\micron)}& \colhead{($j \rightarrow i$)}& \colhead{(s$^{-1}$)}& \colhead{(1000 K)}& \colhead{(5000 K)}& \colhead{($10^{4}$ K)}& \colhead{}& \colhead{(1000 K)}& \colhead{(5000 K)}& \colhead{($10^{4}$ K)}& \colhead{}
}

\startdata

[O III]& 0.4363& $^{1}\mathrm{S}_{0} \rightarrow ~^{1}\mathrm{D}_{2}$& $1.71 \times 10^{+0}$& \nodata& 0.546& 0.647& & \nodata& $1.28 \times 10^{8}$& $1.53 \times 10^{8}$& [1]; NIST\\

& 0.4959& $^{1}\mathrm{D}_{2} \rightarrow ~^{3}\mathrm{P}_{1}$& $6.21 \times 10^{-3}$& \nodata& 0.679& 0.728& & \nodata& $3.75 \times 10^{5}$& $4.94 \times 10^{5}$& [1]; NIST\\

& 0.5007& $^{1}\mathrm{D}_{2} \rightarrow ~^{3}\mathrm{P}_{2}$& $1.81 \times 10^{-2}$& \nodata& 1.131& 1.213& & \nodata& $1.31 \times 10^{5}$& $1.73 \times 10^{5}$& [1]; NIST\\

[O IV]& 25.91& $^{2}\mathrm{P}^{0}_{3/2} \rightarrow ~^{2}\mathrm{P}^{0}_{1/2}$& $5.19 \times 10^{-4}$& 1.641& 2.021& 2.420& & $4.64 \times 10^{3}$& $8.41 \times 10^{3}$& $9.94 \times 10^{3}$& [2]; NIST\\

[Ne II]& 12.81& $^{2}\mathrm{P}^{0}_{1/2} \rightarrow ~^{2}\mathrm{P}^{0}_{3/2}$& $8.59 \times 10^{-3}$& 0.272& 0.277& \nodata& & $4.61 \times 10^{5}$& $1.01 \times 10^{6}$& \nodata& [3]; NIST\\

[Ne III]& 15.56& $^{3}\mathrm{P}_{1} \rightarrow ~^{3}\mathrm{P}_{2}$& $5.84 \times 10^{-3}$& 0.596& 0.739& \nodata& & $1.84 \times 10^{5}$& $3.31 \times 10^{5}$& \nodata& [4]; NIST\\

& 36.01& $^{3}\mathrm{P}_{0} \rightarrow ~^{3}\mathrm{P}_{1}$& $1.10 \times 10^{-3}$& 0.193& 0.226& \nodata& & $2.09 \times 10^{4}$& $3.99 \times 10^{4}$& \nodata& [4]; NIST\\

[Ne V]& 14.32& $^{3}\mathrm{P}_{2} \rightarrow ~^{3}\mathrm{P}_{1}$& $4.59 \times 10^{-3}$& 9.551& 7.653& 5.832& & $8.80 \times 10^{3}$& $2.31 \times 10^{4}$& $4.56 \times 10^{4}$& [5]; ALL\\

& 24.32& $^{3}\mathrm{P}_{1} \rightarrow ~^{3}\mathrm{P}_{0}$& $1.25\times 10^{-3}$& 1.830& 1.688& 1.408& & $7.51 \times 10^{3}$& $1.76 \times 10^{4}$& $3.09 \times 10^{4}$& [5]; ALL\\

[Ne VI]& 7.64& $^{2}\mathrm{P}^{0}_{3/2} \rightarrow ~^{2}\mathrm{P}^{0}_{1/2}$& $2.01\times 10^{-2}$& 2.400& 1.880& 1.600& & $1.23 \times 10^{5}$& $3.50 \times 10^{5}$& $5.82 \times 10^{5}$& [6]; ALL\\

[Mg V]& 5.61& $^{3}\mathrm{P}_{1} \rightarrow ~^{3}\mathrm{P}_{2}$& $8.09 \times 10^{-2}$& 0.556& 0.808& 0.808& & $2.51 \times 10^{6}$& $3.86 \times 10^{6}$& $5.46 \times 10^{6}$& [7]; ALL\\

[Mg VII]& 5.50& $^{3}\mathrm{P}_{2} \rightarrow ~^{3}\mathrm{P}_{1}$& $1.27 \times 10^{-1}$& 0.704& 0.768& 1.079& & $2.11 \times 10^{6}$& $4.32 \times 10^{6}$& $4.34 \times 10^{6}$& [5]; ALL\\

[S IV]& 10.51& $^{2}\mathrm{P}^{0}_{3/2} \rightarrow ~^{2}\mathrm{P}^{0}_{1/2}$& $7.74 \times 10^{-3}$& 5.50& 7.905& 8.536& & $2.06 \times 10^{4}$& $3.21 \times 10^{4}$& $4.20 \times 10^{4}$& [8]; ALL\\

[Ar III]& 8.99& $^{3}\mathrm{P}_{2} \rightarrow ~^{3}\mathrm{P}_{1}$& $3.09 \times 10^{-2}$& 3.88\tablenotemark{c}& 3.84\tablenotemark{c}& \nodata& & $1.46 \times 10^{5}$& $3.30 \times 10^{5}$& \nodata& [9]; ALL \\[-2.5ex]

\enddata

\tablenotetext{a}{The first reference is for $\Upsilon_{ij}$ and the second is for A$_{ij}$. The NIST database is located at http://physics.nist.gov/PhysRefData/ASD/lines\_form.html. The Atomic Line List version 2.05b12 (ALL) is located at http://www.pa.uky.edu/$\sim$peter/newpage/.}
\tablenotetext{b}{REFERENCES: [1] - \citet{Aggarwal83}; [2] - \citet{BlumPradhan92}; [3] - \citet{SaraphTully94}; [4] - \citet{McLaughlinBell00}; [5] - \citet{LennonBurke94}; [6] - \citet{Zhang94}; [7] - \citet{Hudson09}; [8] - \citet{Tayal00}; [9] - \citet{MunozBurgos09}}
\tablenotetext{c}{These values were obtained by linear interpolation of the data reported in the literature.}

\end{deluxetable*}


\section{Analysis}
\label{sec:Analysis}


\subsection{Null Detections}
\label{sec:Null}


\subsubsection{V1500 Cyg (Nova Cygni 1975)}
\label{sec:V1500Cyg}

V1500 Cyg (Nova Cygni 1975) was one of the brightest CNe on record. It was discovered independently by numerous observers around the globe \citep{Honda75} and remains one of the brightest \citep[m$_{\mathrm{V}}$ = 1.9 at max;][]{Young76} and best studied CNe to date. The outburst and decline were the fastest on record with t$_{2}$ = 2.9 days and t$_{3}$ = 3.9 days \citep{HachisuKato06}. The expansion velocity was very high with P-Cyg absorption systems reaching -4200 km s$^{-1}$ and emission line widths of 3000 km s$^{-1}$ FWHM \citep{RosinoTempesti77}. V1500 Cyg was one of the rare novae that transitioned from an \ion{Fe}{2}-type spectrum in the classification system of \citet{Williams92} to that of a He/N-type \citep{DownesDuerbeck00}. The distance was estimated to be 1.5 kpc based upon the expansion parallax of the shell \citep{Slavin95}. No dust was observed in the ejecta. See Table \ref{tab:properties} for a summary of some fundamental characteristics of the system.

We obtained \textit{Spitzer} IRS observations of V1500 Cyg more than 28 years after outburst. No emission lines were observed in these data and the continuum was below our detection limit. Using the weakest detected lines in the high- and low-res modules of V1494 Aql and V1974 Cyg respectively (see \S\ref{sec:V1494Aql_IR} and \S\ref{sec:V1974Cyg_Data}), we estimated the minimum peak flux density above the rms background necessary to clearly detect an emission line. This resulted in scale factors of $4 \times \sigma_{bkgd}$ for the high-res modules and $3 \times \sigma_{bkgd}$ for the emission lines in SL. We then estimated the rms background over the range of $\pm 10000$ km s$^{-1}$ around the lines of interest in V1500 Cyg (i.e. [\ion{Ne}{2}] 12.81, [\ion{Ne}{3}] 15.56, [\ion{Ne}{5}] 14.32 and 24.30, and [\ion{O}{4}] 25.91 \micron) after masking out the line region. Finally, we scaled these background levels by the factors determined for the low- and high-res modules to determine the upper limit to the peak flux densities of the lines. See Table \ref{tab:low-lims}.


\begin{deluxetable*}{lccccc}
\tabletypesize{\scriptsize}
\setlength{\tabcolsep}{0.05in}

\tablewidth{0pt}
\tablecaption{Upper Limits to Line Fluxes for Non-Detected Sources\label{tab:low-lims}}

\tablehead{
\colhead{Object}& \colhead{[\ion{Ne}{2}] 12.81\tablenotemark{a}}& \colhead{[\ion{Ne}{3}] 15.56\tablenotemark{b}}& \colhead{[\ion{Ne}{5}] 14.32\tablenotemark{b}}& \colhead{[\ion{Ne}{5}] 24.30\tablenotemark{b}}& \colhead{[\ion{O}{4}] 25.91\tablenotemark{b}}\\
\colhead{}& \colhead{(Jy)}& \colhead{(Jy)}& \colhead{(Jy)}& \colhead{(Jy)}& \colhead{(Jy)}
}

\startdata

V1500 Cyg& $4.1\times10^{-3}$& $3.5\times10^{-2}$& $2.6\times10^{-2}$& $6.4\times10^{-2}$& $5.1\times10^{-2}$\\
NQ Vul& $1.1\times10^{-2}$& $4.5\times10^{-2}$& $4.9\times10^{-2}$& $8.9\times10^{-2}$& $2.1\times10^{-1}$\\
V1668 Cyg& $3.9\times10^{-3}$& $3.2\times10^{-2}$& $2.4\times10^{-2}$& $5.7\times10^{-2}$& $4.9\times10^{-2}$\\[-2.5ex]

\enddata

\tablenotetext{a}{The upper limits for the low-res module were calculated assuming $3 \times \sigma_{bkgd}$}
\tablenotetext{b}{The upper limits for the high-res modules were calculated assuming $4 \times \sigma_{bkgd}$}

\end{deluxetable*}



\subsubsection{NQ Vul (Nova Vulpeculae 1976)}
\label{sec:NQVul}

NQ Vul (Nova Vulpeculae 1976) was discovered by G.\ E.\ D.\ Alcock on 1976 Oct 21.7 UT \citep{Milbourn76} at a brightness of m$_{\mathrm{V}}$ = 6.5. After detection, the system underwent a rebrightening event that peaked on Nov 3 at a visual magnitude of 6.0 \citep{DuerbeckSeitter79}. Characterization of the t$_{2}$ and t$_{3}$ times were difficult because of the erratic nature of the light curve, but t$_{3}$ was estimated to be $\sim 53$ days \citep{Yamashita77}. The nova exhibited expansion velocities of order 705 km s$^{-1}$ \citep{CohenRosenthal83} and the spectra were typical of an \ion{Fe}{2} nova \citep{DownesDuerbeck00}. \citet{DownesDuerbeck00} estimated a distance to NQ Vul of 1.2 kpc by averaging the distances reported in the literature. About a month after outburst, CO formed in the ejecta, the first detection of that molecule in a CN \citep{Ferland79}. This preceded the formation of carbon-rich dust that exhibited only a weak extinction event in the light curve around day 70 \citep{NeyHatfield78}.

The \textit{Spitzer} data on NQ Vul were obtained at roughly 27.5 years after outburst. Like V1500 Cyg, the spectra were  featureless with negligible continuum. Upper limits to the flux densities of important forbidden lines were estimated using the same methodology as described above (\S\ref{sec:V1500Cyg}). The results are provided in Table \ref{tab:low-lims}.


\subsubsection{V1668 Cyg (Nova Cygni 1978)}
\label{sec:V1668Cyg}

V1668 Cyg (Nova Cygni 1978) was discovered on 10.24 Sept. 1978 at m$_{\mathrm{V}} \sim 7.0$ \citep{Collins78}, just two days before maximum, which reached m$_{\mathrm{V}} = 6.0$ \citep{HachisuKato06}. \citet{HachisuKato06} classified the system as a fast nova based on the t$_{2}$ time of 12.2 days (see Table \ref{tab:properties}) and suggested that the eruption arose on a CO WD, based upon their model light curves. A CO progenitor is in general agreement with abundance estimates by \citet{Stickland81}, which showed the most significant enrichment in CNO processed materials. A shell of optically thin dust began to condense around 30 days after outburst.

\textit{Spitzer} observations of V1668 Cyg were made approximately 25 years after outburst and failed to detect either significant continuum or any indication of emission lines. Again, upper limits on the flux densities of important nebular lines were estimated using the methods described in \S\ref{sec:V1500Cyg} and are presented in Table \ref{tab:low-lims}.


\subsection{V1974 Cyg (Nova Cygni 1992)}
\label{sec:V1974Cyg}

V1974 Cyg (Nova Cygni 1992) was discovered before maximum light by \cite{Collins92} on 1992 February 19.1UT at a magnitude of m$_{V}=6.8$. The nova reached maximum brightness on 1992 February 20.8UT (JD 2448673.3), which we take to be t$_{0}$, at m$_{V}$=4.3 \citep{Schmeer92}. The exceedingly bright outburst of V1974 Cyg allowed what was for the time an unprecedented degree of coverage. Initially, the light curve showed slight variations that complicated the determination of t$_{2}$, the time it took the V-band light curve to decay by two magnitudes, which is often used in conjunction with standard Maximum-Magnitude-Rate of Decline (MMRD) relations to determine the distance to the target. Estimates of t$_{2}$ ranged from 16 to 24 days \citep{Chochol97, Austin96, Quirrenbach93}. These estimates situate the nova in the ``fast'' category of CNe \citep{Gaposchkin57}. The fundamental properties of V1974 Cyg are summarized in Table \ref{tab:properties}.

The detection of \ion{Fe}{2} lines, with accompanying permitted emission from species such as \ion{Ca}{2}, \ion{Mg}{2}, \ion{N}{2}, \ion{Na}{1}, and \ion{O}{1} \citep{Baruffolo92,Chochol93}, revealed this to be an ``Fe II''-type nova. A multicomponent P-Cygni profile was observed in H$\beta$ very early in the outburst with a peak velocity of -1670 km s$^{-1}$ \citep{Garnavich92}. Later analysis of the absorption components revealed that they accelerated with a terminal velocity of v$_{\infty} = 2899 \pm 8$ km s$^{-1}$ \citep{Cassatella04}. \citet{Gehrz92a} detected emission from [\ion{Ne}{2}] 12.81 \micron ~just over a week after outburst and subsequent observations of strong neon emission in the optical prompted \citet{Austin96} to classify V1974 Cyg as a neon nova. [\ion{Ne}{2}] and other mid-IR lines exhibited HWHM velocities of $\sim$ 1500 km s$^{-1}$ \citep{Gehrz92b, Woodward95}, comparable to the initial velocity of the P-Cygni absorption system.

V1974 Cyg transitioned into the coronal stage of nebular development within $\sim 150$ days of outburst with the emergence of [\ion{Ne}{5}] $\lambda\lambda$3346,3426 \AA\ \citep{Barger93}. This was followed by the detection of numerous coronal lines in the IR, including [\ion{Al}{6}] 3.661 \micron, [\ion{Al}{8}] 3.720 \micron, [\ion{S}{9}] 1.250 \micron, [\ion{Mg}{8}] 3.028 \micron, and [\ion{Ca}{9}] 3.088 \micron\ \citep{Woodward95}. There was no evidence for significant dust production in the ejecta.

\subsubsection{V1974 Cyg - \textit{Spitzer} Data}
\label{sec:V1974Cyg_Data}

\begin{figure}
\epsscale{1.1}
\plotone{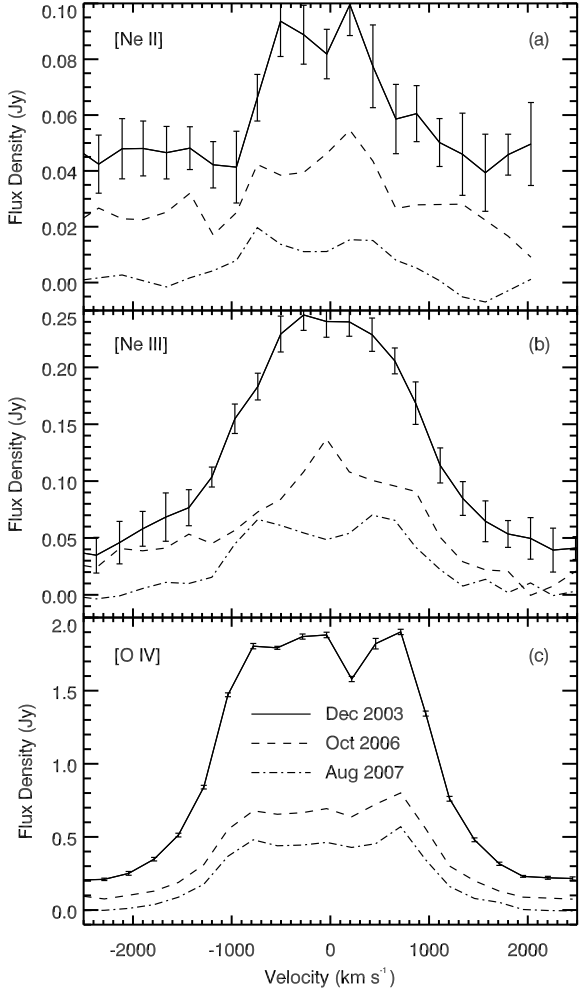}
\caption{V1974 Cyg: Emission lines observed at three different epochs. Representative errors are shown on the first epoch data and are comparable for the next two epochs. \textit{Panel (a)}: Emission from [\ion{Ne}{2}] 12.81 \micron\ is very weak. The profile is dominated by the noise in the data. \textit{Panel (b)}: The [\ion{Ne}{3}] 15.56 \micron\ emission is relatively strong. The Dec. 2003 profile shape is real, but the profile in the later two epochs are much more sensitive to the noise level. \textit{Panel (c)}: The [\ion{O}{4}] 25.91 \micron\ line was the strongest in the spectra. The profile shape is real, and the consistent shape between epochs indicates that the ejecta distribution was no longer subjected to shaping effects at these late times. \label{fig:V1974Cyg_MIR}}
\end{figure}

Our IR spectra were obtained 11.8, 14.7, and 15.5 years after outburst. They showed emission only from [\ion{Ne}{2}] 12.81 \micron, [\ion{Ne}{3}] 15.56 \micron, and [\ion{O}{4}] 25.91 \micron, with negligible continuum. The profiles of these lines are presented in Figure \ref{fig:V1974Cyg_MIR}. The signal-to-noise (S/N) for all of the observations of [\ion{Ne}{2}] and the last two epochs of [\ion{Ne}{3}] was very low (S/N $\simeq 2$), making accurate determination of flux values and characterization of the profiles difficult. On the other hand, the S/N of the first observation of [\ion{Ne}{3}] and all three observations of [\ion{O}{4}] was high, bearing in mind that the formal errors are somewhat overestimated (see \S\ref{sec:OldNovae_MIR}). A summary of the line measurements are provided in Table \ref{tab:V1974Cyg_Lines}.


\begin{deluxetable*}{lcccccccccc}
\tabletypesize{\scriptsize}
\setlength{\tabcolsep}{0.05in}

\tablewidth{0pt}
\tablecaption{V1974 Cyg - Line Measurements \label{tab:V1974Cyg_Lines}}

\tablehead{
\colhead{}& \colhead{$\lambda_{o}$}& & \multicolumn{2}{c}{2003 Dec. 17.4}& & \multicolumn{2}{c}{2006 Oct. 18.4}& & \multicolumn{2}{c}{2007 Aug. 30.2} \\[0.5ex]
\cline{4-5} \cline{7-8} \cline{10-11}\\[-1.8ex]
\colhead{Ion}& \colhead{(\micron)}& Module& Flux\tablenotemark{a}& FWHM\tablenotemark{b}& & Flux\tablenotemark{a}& FWHM\tablenotemark{b}& & Flux\tablenotemark{a}& FWHM\tablenotemark{b} 
}

\startdata

$[$Ne II$]$& 12.81& SL,SH\tablenotemark{c}& $0.5 \pm 0.26$& $1200 \pm 400$& & $0.3 \pm 0.4$& $1400 \pm 1200$& & $0.2 \pm 0.1$& $1700 \pm 600$ \\
$[$Ne III$]$& 15.56& SH& $2.8 \pm 0.3$& $1900 \pm 200$& & $1.1 \pm 0.3$& $1600 \pm 400$& & $0.9 \pm 0.1$& $2090 \pm 140$ \\
$[$O IV$]$& 25.91& LH& $14.9 \pm 0.6$& $2350 \pm 50$& & $5.8 \pm 0.4$& $2360 \pm 100$& & $4.5 \pm 0.3$& $2320 \pm 80$\\[-2.5ex]

\enddata

\tablenotetext{a}{Fluxes provided in units of $10^{-13}$ erg s$^{-1}$ cm$^{-2}$}
\tablenotetext{b}{FWHM velocity widths in km s$^{-1}$, uncorrected for the instrument resolution of $\sim 500$ km s$^{-1}$.}
\tablenotetext{c}{When both high- and low-resolution data were available, we preferentially selected the high-res data for measurement of line parameters.}

\end{deluxetable*}


\subsubsection{V1974 Cyg - Nebular Environment}
\label{sec:V1974Cyg_Line_Struct}

The resolution of the \textit{Spitzer} SH and LH modules is $R \approx 600$, yielding a resolution element of $\sim 500$ km s$^{-1}$. The emission lines detected by \textit{Spitzer} have FWHM velocity widths between $\sim 2000$ and 2300 km s$^{-1}$, which indicates that they are resolved and the structure seen in the profiles of the lines with high S/N is real. There are distinct differences between the rounded shape of the low ionization [\ion{Ne}{3}] emission profile and the flat-topped profile of the more highly ionized [\ion{O}{4}] line. \citet{Salama96} noted the same trend for the [\ion{Ne}{3}], [\ion{O}{4}], and [\ion{Ne}{5}] lines in their Infrared Space Observatory (ISO) observations of V1974 Cyg taken 1494 days post-outburst. At that time, the [\ion{Ne}{3}] 15.56 \micron\ and [\ion{O}{4}] 25.91 \micron\ lines (with ionization potentials, I.P., of 41.0 and 63.5 eV respectively) both exhibited relatively simple, rounded profiles while the [\ion{Ne}{5}] 14.32 and 24.30 \micron\ lines (I.P. = 97.1 eV) had structured, saddle-shaped profiles. Salama et al.\ suggested that the structure seen in the line profiles might indicate that the ejecta had an equatorial ring / polar cap morphology with additional tropical rings. 

The contrast between the ISO line profiles obtained earlier in the development of V1974 Cyg and the later \textit{Spitzer} profiles is interesting. Whereas in the ISO observations the [\ion{O}{4}] line and the low ionization [\ion{Ne}{3}] line were rounded, in the \textit{Spitzer} data the [\ion{O}{4}] line had transitioned to a flat-topped profile much like the ISO observations of the more highly ionized neon lines. The simplest geometry sufficient to explain the rounded profiles is that of an optically thin, geometrically thick shell. Similarly the flat topped profiles can be described by an optically and geometrically thin shell. The transition in the [\ion{O}{4}] profile shape suggests that as the system aged, the degree of ionization in the kinematic components identified by Salama et al.\ declined and the component of the ejecta providing the dominant contribution to the emission line shifted from the geometrically thick shell to the thin shell. 

The widths of the late time \textit{Spitzer} [\ion{O}{4}] line are consistently $\sim 200$ km s$^{-1}$ wider than that measured by Salama et al. This may suggest that the rounded profile of the ISO observation of [\ion{O}{4}] was a superposition of two components, one relatively weak, flat-topped and broad, and one slightly narrower and rounded. As the system evolved, the round component faded more dramatically than the flat-topped component, behavior consistent with a more rapid decline in ionization in the interior, geometrically thick component of the ejecta. The differences between the profile shapes of the O and Ne lines, may be due to a different shell depths and density distributions for the emitting species, with the [\ion{Ne}{3}] emission arising from a region of the shell with greater thickness than that from which the [\ion{O}{4}] arises. 

\begin{figure}
\epsscale{1.2}
\plotone{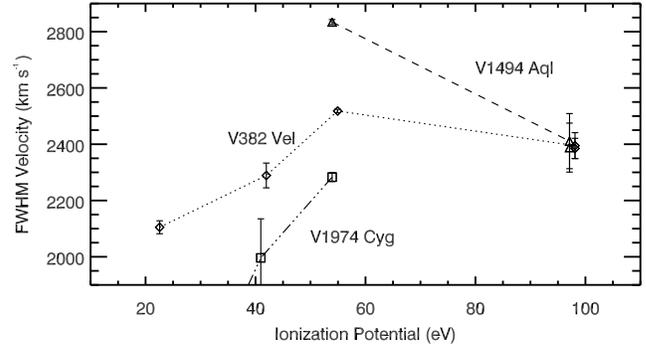}
\caption{This plot shows the FWHM velocity of the emission lines measured from the first epoch of \textit{Spitzer} observations for each of the targets as a function of the I.P. necessary to produce the ion giving rise to the observed emission line. The data points for V382 Vel are offset by +1 eV for clarity.  \label{fig:FWHMEvolve}}
\end{figure}

Additional clues to the structure of the ejecta may be derived from the differences in the emission line widths. Figure \ref{fig:FWHMEvolve} shows a plot of the FWHM velocity of the [\ion{Ne}{3}] 15.56 \micron\ and the [\ion{O}{4}] 25.91 \micron\ lines as a function of I.P. The error on the velocity width of the [\ion{Ne}{2}] 12.81 \micron\ line is high enough to be considered unreliable. Based on these data and under the standard assumption that the velocity increases outward in the ejecta, one might expect that the [\ion{O}{4}] emission, which arises from higher velocity material, originates from a region exterior to the [\ion{Ne}{3}] emission. 

\begin{figure}
\epsscale{1.2}
\plotone{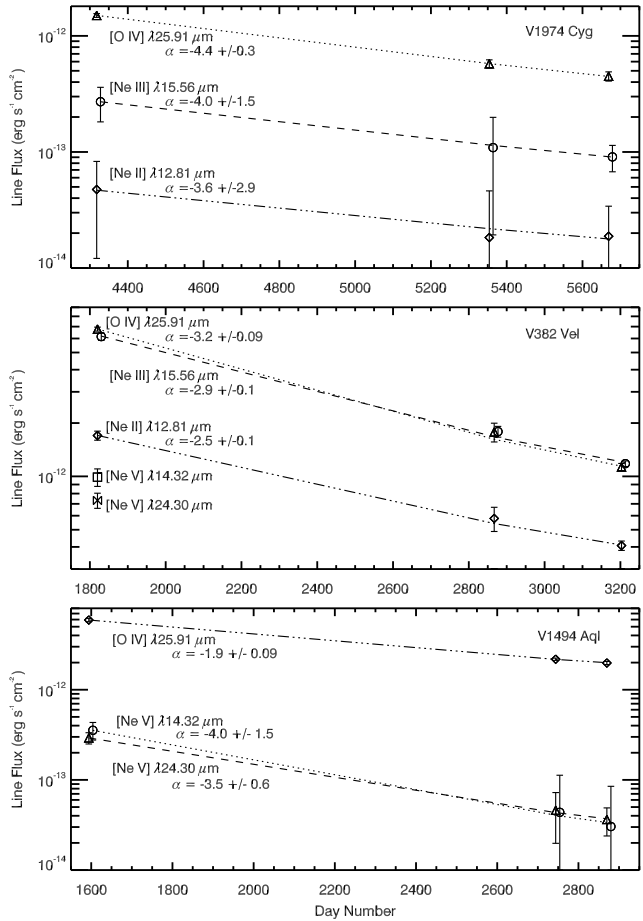}
\caption{This figure presents the evolution of the integrated line flux (erg s$^{-1}$ cm$^{-2}$) for each of the three targets. The dashed, dotted, and dash-dotted lines represent the best fit exponential decay of the form F $\propto t^{\alpha}$, with the decay parameter $\alpha$ given. \textit{Top}: V1974 Cyg; \textit{Middle}: V382 Vel; \textit{Bottom}: V1494 Aql. \label{fig:FluxEvolve}}
\end{figure}

If the hypothesis of two kinematically distinct components in the ejecta is correct, then one might expect that the rate of decline in line strength of each component would behave differently. For example, if the material is distributed in a uniformly expanding, filled sphere, then the expected decline in line flux would be $\propto t^{-\alpha}$, where $\alpha = 6$, since the flux is dependent on both the geometrical dilution of the ejecta and the square of the density. If the emitting material is distributed in a thin shell with a fixed shell depth, $\Delta$r, then one would expect F $\propto t^{-4}$. In the top panel of Figure \ref{fig:FluxEvolve}, we have plotted the evolution of the line fluxes measured in V1974 Cyg and have fitted the data with a function of the form $F(t) = a \cdot t^{-\alpha}$, where $a$ is a simple scaling parameter, using a non-linear least squares Marquardt fitting algorithm. Within the errors, the decline rates of both the neon and the oxygen lines are consistent with $\alpha \simeq 4$, as might be expected for a thin shell model. 

Direct imaging of the shell of V1974 Cyg has been undertaken in the optical and near-IR by \citet{Paresce93,Paresce94,DownesDuerbeck00,Krautter02}. Krautter et al.\ used the Hubble Space Telescope and Downs \& Duerbeck used multiple ground-based optical telescopes. The images from these studies suggest a slightly ellipsoidal, limb-brightened shell with an inhomogeneous mass distribution. The IR emission lines presented here appear to arise from a similar distribution of material. \citet{Krautter02} also noted the possible detection of an additional ring component along with the ellipsoidal shell, though this component does not appear to be reflected in our \textit{Spitzer} data.

\subsubsection{V1974 Cyg - Adopted Parameters}
\label{sec:v1974_vals}

Several authors have estimated the distance to V1974 Cyg using a variety of methods. \citet{Chochol97} calculated a weighted mean distance of 1.77 $\pm$ 0.11 kpc, which relied heavily on various forms of the MMRD relation \citep[]{DVL95}. Though often used, the MMRD may be unreliable for determining distances to ONe novae \citep{DelPozzo05}. Other distance estimates using more reliable independent measurements of the expansion parallax were not included in the work of \citet{Chochol97}. In Table \ref{tab:distances}, we present these distance estimates and calculate a weighted mean distance of $3.0 \pm 0.1$ kpc, which we use in subsequent analysis. 


\begin{deluxetable*}{llcl}
\tabletypesize{\scriptsize}
\setlength{\tabcolsep}{0.05in}

\tablewidth{0pt}
\tablecaption{Distance Estimates \label{tab:distances}}

\tablehead{
\colhead{Source}& \colhead{Reference}& \colhead{Distance (kpc)}& \colhead{Method}
}

\startdata

V1974 Cyg& \cite{Cassatella04}& $2.9 \pm 0.2$& Expansion parallax based on\\
& & & ~~~\cite{Quirrenbach93}\\
& \cite{Krautter02}& $3.72 \pm 0.55$\tablenotemark{a}& Expansion parallax\\
& \cite{Shore94}& 2.3 - 4.6& Expansion parallax\\[0.5ex]
\cline{2-4}\\[-1.5ex]
& \textit{Weighted Mean}& $3.0 \pm 0.2$& \\[1ex]
\hline\\[-1.5ex]
V382 Vel& \cite{DellaValle02}& $\sim 1.7$& MMRD relation \\
& \cite{Shore03}& 2& MMRD relation \\
& & $>2$& $^{12}$CO absorption \\
& & 2 - 3& Comparison to Nova LMC 2000 \\
& & $> 2$& Interstellar absorption lines \\
& & 2& Pre-outburst luminosity \\
& & 2.5& L$_{max} \leq$~L$_{Edd}$ for M$_{WD} = 1.4$~M$_{\odot}$ \\
& & 2.5& Photoionization models \\[0.5ex]
\cline{2-4}\\[-1.5ex]
& \textit{Chosen Value\tablenotemark{b}}& 2.3& \\[1ex]
\hline\\[-1.5ex]
V1494 Aql& \cite{KissThomson00}& $3.6 \pm 0.3$& MMRD relation \\
& \cite{IijimaEsenoglu03}& $1.6 \pm 0.2$& Na I D interstellar absorption \\
& \cite{Hachisu04}& 1.0 - 1.4& Light curve model \\[0.5ex]
\cline{2-4}\\[-1.5ex]
& \textit{Chosen Value\tablenotemark{b}}& 1.6& \\[-2.5ex]
\enddata

\tablenotetext{a}{The error used in the weighted mean calculation is based upon the spread in adopted values of v$_{exp}$.}
\tablenotetext{b}{The distance value was selected on qualitative grounds since there were few reliable quantitative estimates.}
\end{deluxetable*}

The mass of ejecta generated in the outburst of V1974 Cyg has been estimated by a number of techniques. \citet{Shore93} used early spectroscopic observations of He II $\lambda$1640 \AA\ to determine the helium enrichment dependent ejecta mass. Using the relationship derived by Shore et al.\ and the He abundance from their Cloudy photoionization models, \citet{Vanlandingham05} estimated an ejecta mass of M$_{ej} = 1.9 \times 10^{-4}$ M$_{\odot}$. This estimate was reinforced by calculations of the mass of ejecta directly from their Cloudy models for three different dates, which yielded a weighted mean M$_{ej} = (2.2 \pm 0.1) \times 10^{-4}$ M$_{\odot}$. We adopt $2 \times 10^{-4}$ M$_{\odot}$ for our analysis.

NIR observations of the emission lines observed throughout the first $\sim500$ days of the evolution in the near-IR indicated that the emission line widths remained nearly constant at 2500 -- 3000 km s$^{-1}$ \citep{Woodward95}. The mid-IR neon lines observed with \textit{Spitzer} and presented here exhibit a similar width. Hence, we take the minimum expansion velocity to be v$_{min} = 1250$ km s$^{-1}$. Assuming that the maximum velocity observed in the diffuse enhanced absorption system represents the fastest moving material, we take v$_{max}$ to be 2900 km s$^{-1}$. 

We calculated the electron density assuming two different geometries, both of which were spherically symmetric. The first consisted of a spherical shell expanding at 2900 km s$^{-1}$ with a shell depth of $\Delta r = 0.1r_{out}$, where $r_{out}$ is the radius of the outer boundary of the ejecta. The other was a spherical shell with v$_{min} = 1250$ km s$^{-1}$ and v$_{max} = 2900$ km s$^{-1}$. These geometries yielded electron densities, $n_{e}$, of $\sim 50$ cm$^{-3}$ and $\sim 170$ cm$^{-3}$ respectively.

Attempts to constrain the electron temperature using the [\ion{O}{3}] doublet at $\lambda\lambda$4959, 5007 \AA\ relative to [\ion{O}{3}] $\lambda$4363 \AA\ failed because at these late epochs, the 4363 \AA\ transition was undetectable while the doublet was only weakly detected. Thus, we were forced to adopt temperatures based upon earlier measurements. Early after outburst (day 23.5), \citet{Woodward95} estimated the electron temperature to be $T_{e} \sim 5000$ K declining to $\sim 1000$~K by day 180. This low temperature is consistent with expectations for nebular material that is enriched in metals and undergoing efficient cooling by IR coronal lines \citep{Woodward97,Smits91}. Using the line fluxes of Paschen-$\alpha$ and Brackett-$\gamma$ obtained by \citet{Krautter02}, we calculate a line ratio of [\ion{H}{1} (7-4) / \ion{H}{1} (4-3)] = $(4.90 \pm 0.57) \times 10^{-2}$ (uncorrected for reddening), suggesting that T$_{e} \leq 1000$~K, with n$_{e} \ltsimeq 1000$~cm$^{-3}$ \citep[assuming Case B recombination;][]{HummerStorey87}. We chose to calculate the abundances using $1000 <$~$T_{e}$~(K) $< 5000$; this range will allow us to assess the temperature dependence of the derived abundances, though we emphasize that the temperature is likely closer to the lower value.

A summary of the adopted parameters for V1974 Cyg along with the targets that follow is provided in Table \ref{tab:parameters}.


\begin{deluxetable}{lccc}
\tabletypesize{\scriptsize}
\setlength{\tabcolsep}{0.05in}

\tablewidth{0pt}
\tablecaption{Adopted Parameters \label{tab:parameters}}

\tablehead{
\colhead{Parameter}& \colhead{V1974 Cyg}& \colhead{V382 Vel}& \colhead{V1494 Aql}
}

\startdata

D (kpc)& 3.0& 2.3& 1.6\\
M$_{ej}$ (M$_{\odot}$)& $2 \times 10^{-4}$& $4 \times 10^{-4}$& $2.5 \times 10^{-5}$\\
v$_{0}$ (km s$^{-1}$)\tablenotemark{a}& 2900& 3500& 1850\\
v$_{0}$ (km s$^{-1}$)\tablenotemark{b}& 1250-2900& 1200-3500& 980-2500\\
$T_{e}$ (K)& 1000-5000& 1000-5000& $10^{4}$\\
$n_{e}$ (cm$^{-3}$)& 50 -- 170& 700 -- 2500& 650 -- 1450\\
n$_{e,clump}$ (cm$^{-3}$)& \nodata& \nodata& $10^{6}$ -- $10^{7}$\\[-2.5ex]

\enddata

\tablenotetext{a}{This is the value used to calculate the volume of a spherical shell assuming $\Delta r / r \sim 0.1$.}
\tablenotetext{b}{These are the values used to calculate the shell volume for other geometries as described in the text.}

\end{deluxetable}


\subsubsection{V1974 Cyg - Abundances}
\label{sec:v1974_Ab}

The results of our abundance analysis using the parameters described in \S\ref{sec:v1974_vals} are presented in Table \ref{tab:v1974_ab}. At T$_{e} = 1000$~K, the minimum Ne ionic abundance by number was 41 times solar, while a temperature of 5000 K resulted in a relative abundance only marginally lower at 35 times solar. The difference between these estimates is insignificant relative to the errors intrinsic to the calculations. The oxygen abundance was found to be more than 9 times solar. Relative to O$^{3+}$, (Ne$^{2+}$ + Ne$^{3+}$) was found to be $\gtsimeq 3.9$~times solar. The shell geometries used in each calculation dominated the uncertainty in the abundance estimates. In the low density model ($n_{e} \sim 50$ cm$^{-3}$), the derived abundances were higher than the high density model ($n_{e} \sim 170$ cm$^{-3}$) by a factor of $\gtsimeq 3$, and so, the abundance estimation was clearly more sensitive to the assumed $n_{e}$ than $T_{e}$.


\begin{deluxetable*}{lccccccc}
\tabletypesize{\scriptsize}
\setlength{\tabcolsep}{0.025in}

\tablewidth{0pt}
\tablecaption{Abundances by Number Relative to Solar - V1974 Cyg \label{tab:v1974_ab}}

\tablehead{
\colhead{}& \colhead{}& \colhead{}& \multicolumn{5}{c}{Derived Abundances\tablenotemark{a}}\\[0.5ex]
\cline{4-8}\\[-1.8ex]
\colhead{Species}& \colhead{Wavelength}& \colhead{$n_{e}$ ($cm^{-3}$):}& \multicolumn{2}{c}{50}& \colhead{}& \multicolumn{2}{c}{170}\\
\cline{4-5} \cline{7-8}\\[-1.5ex]
\colhead{}& \colhead{(\micron)}& \colhead{$T_{e}$ (K):}&
\colhead{ \ 1000 \ }& \colhead{ \ 5000 \ }& \colhead{}& \colhead{ \ 1000 \ }& \colhead{ \ 5000 \ }
}

\startdata

[Ne II]& 12.81& & -2.64& -2.69& & -3.17& -3.22\\

[Ne III]& 15.56& & -2.02& -2.09& & -2.56& -2.62\\[0.5ex]

\hline\\[-1.5ex]

\textbf{Total Neon}& & & -1.93 (138)& -1.99 (119)& & -2.46 (40.6)& -2.52 (35.2)\\[0.5ex]

\hline\\[-1.5ex]

[O IV]& 25.91& & -1.84 (29.2)& -1.78 (33.8)& & -2.35 (9.1)& -2.29 (10.5)\\[-2.5ex]

\enddata

\tablenotetext{a}{The number given is the log abundance by number relative to hydrogen. The number in parentheses is the abundance by number relative to hydrogen, relative to solar assuming the solar values of \citet{Asplund09}.}

\end{deluxetable*}


The abundance estimates derived here for Ne and O using the late-time \textit{Spitzer} spectra are consistent with those derived earlier in the outburst by other authors. In particular, the photoionization models of \citet{Vanlandingham05} indicated that V1974 Cyg was overabundant in oxygen relative to solar by a factor of 12.8 $\pm$ 7 and neon by 41.5 $\pm$ 17 while IR observations of [\ion{Ne}{3}] and [\ion{O}{4}] by \citet{Salama96} revealed a Ne to O ratio of $\sim 4$ times solar. These global abundance solutions were very similar to the abundances derived for individual knots in the ejecta by \citet{Paresce95} who found overabundances of 20-30 and 2-4 times solar for Ne and O respectively. Likewise, \citet{Shore97} estimated the abundances in a single knot to be [O/H]=6.9 and [Ne/H]=15.6. 

The most discrepant reports were by \citet{Gehrz94}, who estimated neon to be about 10 times the solar value. However, we expect the uncertainty in this last estimate to be high since it relied on only two lines, [\ion{Ne}{6}] 7.62 \micron ~and [\ion{Si}{6}]  1.959 \micron, and assumed that the relative abundance of Ne to Si derived from those lines could be used as a proxy for the Ne abundance relative to solar.

\subsection{V382 Vel (Nova Velorum 1999)}
\label{sec:V382Vel}

V382 Vel (Nova Velorum 1999) was discovered by P. Williams and independently by Gilmore on 1999 May 22.4UT (JD 2451320.9) at a magnitude of m$_{V}=3.1$ \citep{Lee99} shortly before maximum on May 23 (t$_{0}$) near m$_{V} \sim 2.3$ \citep{DellaValle02}. The light curve decay was very rapid and smooth with decay times of t$_{2}=(4.5-6)$ and t$_{3}=(9-12.5)$ \citep{DellaValle99,Liller00} suggesting that V382 Vel was a ``very fast'' nova \citep{DellaValle02,Gaposchkin57}. A summary of the photometric characteristics is provided by \citet{DellaValle02}. Basic system properties are provided in Table \ref{tab:properties}.

Early optical spectra \citep{DellaValle02} revealed a complex spectrum with prominent permitted \ion{Fe}{2} emission and hydrogen Balmer lines. Though additional emission from \ion{Na}{1}, \ion{Al}{2}, \ion{Mg}{2}, and \ion{Ca}{2} were present, the strongest lines were from \ion{O}{1} $\lambda$7773 \AA\ and \ion{O}{1} $\lambda$8446 \AA. The emission lines displayed strong P-Cyg absorption components at terminal velocities of -2300 and -3700 km s$^{-1}$, characterizing the ``Principal'' and ``Diffuse Enhanced'' absorption systems, respectively \citep{DellaValle02}. The emission lines were broad (HWZI $\gtsimeq 3500$ km s$^{-1}$) and relatively flat-topped, suggesting that they originated in a discrete shell. The broad line widths of the iron lines indicated that this was an ``\ion{Fe}{2}b'' (broad) type CN \citep{Steiner99,DellaValle99}. 

Within a month, the system transitioned into the nebular stage with the appearance of [\ion{O}{3}] $\lambda\lambda$4959, 5007 \AA, \ion{He}{1} $\lambda$4471, 6678, and 7065 \AA, \ion{He}{2} $\lambda$5876 \AA, and [\ion{N}{2}] $\lambda$5755 \AA\ \citep{DellaValle02}. During the early nebular stage, the UV emission was still within the ``iron curtain'' stage and consequently optically thick. As this stage faded, P-Cyg absorption components remained in evidence on the resonance lines with absorption maxima near 5000 km s$^{-1}$. This behavior is consistent with the behavior of other ONe novae such as V1974 Cyg \citep{Shore03}. The classification of V382 Vel as an ONe nova was based on the detection of prominent [\ion{Ne}{2}] 12.81 \micron\ emission at 43.6 days after maximum \citep{Woodward99} and was reinforced by the detection of exceptionally strong emission from [\ion{Ne}{3}] $\lambda\lambda$3869, 3968 \AA\ at nearly the same time \citep{AD03}. At this stage of development, there was still emission evident from \ion{H}{1} (9-7) and (7-6) with a FWHM velocity of 3300 km s$^{-1}$ for \ion{H}{1} (7-6) \citep{Woodward99}. 

Late in the nebular stage, the overall state of ionization in the ejecta increased. The optical spectra exhibited emission from [\ion{Ne}{4}] $\lambda$4721 \AA, [\ion{Ne}{5}] $\lambda\lambda$3346, 3426 \AA, additional lines of \ion{He}{2} at 5412 and 7593 \AA, and strong emission from [\ion{Fe}{6}] $\lambda$ 5677 \AA\ and [\ion{Fe}{7}] at 3759, 5158, 5276, and 6087 \AA\ \citep{DellaValle02, AD03}. As the ejecta evolved, they also showed emission from [\ion{Fe}{10}] $\lambda$6375 \AA\ \citep{DellaValle02}, which is commonly seen in novae during their super-soft source stage (SSS) of X-ray development \citep{Ness07,Schwarz11}. Indeed, Beppo-SAX observations of V382 Vel taken during this same period found that it had transitioned from an early stage of hard X-ray emission into a relatively strong SSS \citep{Orio99a,Orio99b}. Later \textit{Chandra} observations showed that the SSS spectrum decayed quite rapidly between 222 and 268 days after outburst, indicating that the central source had turned off \citep{Burwitz02}. There was no evidence for dust production.

\subsubsection{V328 Vel - \textit{Spitzer} Data}
\label{sec:V382Vel_Data}

\begin{figure}
\epsscale{1.2}
\plotone{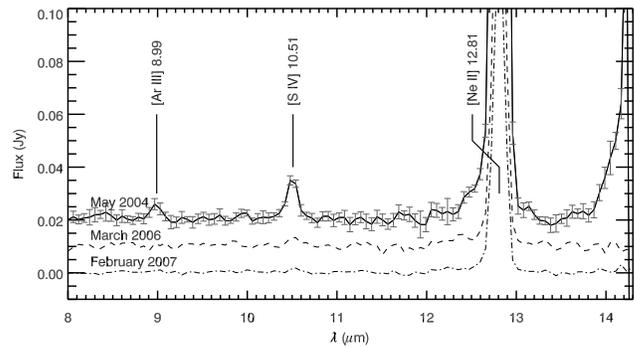}
\caption{V382 Vel: \textit{Spitzer} SL spectra of V382 Vel obtained on 2004 May 14, 2007 March 26, and 2008 February 25. Prominent emission lines are labeled. The strong feature beyond 14 \micron\ in the 2004 May spectrum is due to [\ion{Ne}{5}] 14.32. The shape of this feature in the spectrum is not well characterized due to its proximity to the end of the spectral range of the SL module.\label{fig:V382Vel_Low}}
\end{figure}

\begin{figure}
\epsscale{1.2}
\plotone{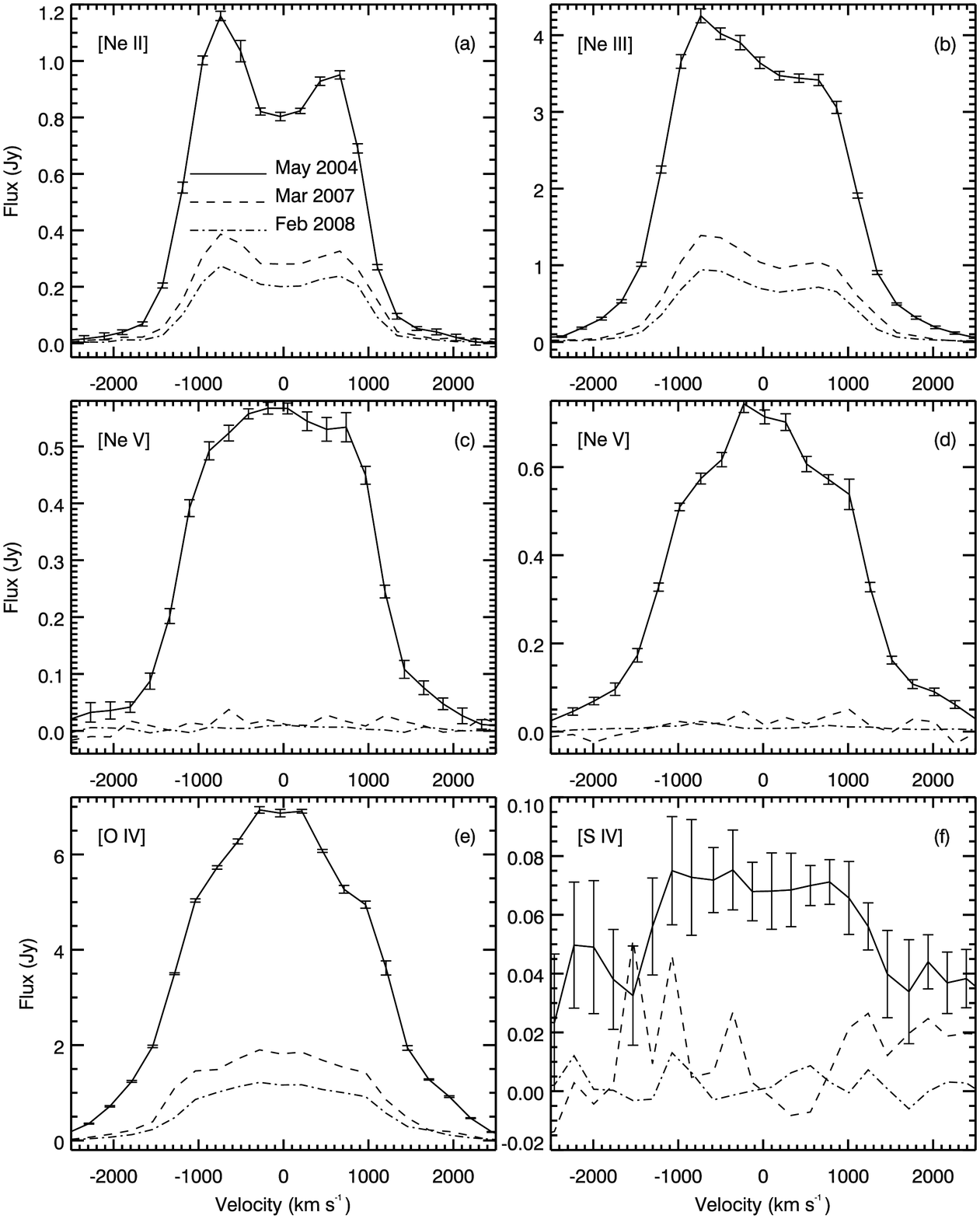}
\caption{V382 Vel: Emission lines observed at three different epochs with the \textit{Spitzer} high resolution modules. Representative errors are shown on the first epoch data and are comparable for the next two epochs. \textit{Panel (a)}: Emission from [\ion{Ne}{2}] 12.81 \micron\ is relatively strong and persists throughout all three epochs. The profile has a saddle shape that is consistent throughout our observations. \textit{Panel (b)}: The [\ion{Ne}{3}] 15.56 \micron\ emission is exceptionally strong. The saddle-shaped structure is less well defined, but is similar to the [\ion{Ne}{2}] line in that it is dominated by a blue-ward peak. \textit{Panels (c-d)}: The [\ion{Ne}{5}] lines at 14.32 and 24.30 \micron\ were strong only during the first epoch of observations after which they were no longer detected. Their line profiles are rounded, distinctly different than the [\ion{Ne}{2}] and [\ion{Ne}{3}] lines. \textit{Panel (e)}: The [\ion{O}{4}] 25.91 \micron\ line was the strongest in the spectra. The profile shape is rounded like the [\ion{Ne}{5}] lines, which suggests that these features arose in the same regions of the ejecta. \textit{Panel (f)}: The [\ion{S}{4}] 10.51 \micron\ line was barely detected above the noise. Due to the low signal-to-noise in this line, its line profile is not well defined.\label{fig:V382Vel_MIR}}
\end{figure}

The \textit{Spitzer} data were obtained 8.6, 11.5, and 12.4 years after outburst (Tab. \ref{tab:sp_obs}) and are presented in Figures \ref{fig:V382Vel_Low} and \ref{fig:V382Vel_MIR}. The first epoch spectra were dominated by [\ion{Ne}{2}] 12.81 \micron, [\ion{Ne}{3}] 15.56 \micron, and [\ion{O}{4}] 25.91 \micron. Emission lines of [\ion{Ne}{5}] were present at 14.32 \micron\ and 24.30 \micron, but at a much lower level than the other neon species. No [\ion{Ne}{6}] 7.61 \micron\ emission was detected. There was also weak emission from [\ion{S}{4}] 10.51 \micron\ and [\ion{Ar}{3}] 8.99 \micron. Subsequent observations detected only emission from [\ion{Ne}{2}], [\ion{Ne}{3}], and [\ion{O}{4}]. A summary of the line measurements are provided in Table \ref{tab:V382Vel_Lines}.

\begin{deluxetable*}{lcccccccccc}
\tabletypesize{\scriptsize}
\setlength{\tabcolsep}{0.05in}

\tablewidth{0pt}
\tablecaption{V382 Vel - Line Measurements \label{tab:V382Vel_Lines}}

\tablehead{
\colhead{}& \colhead{$\lambda_{o}$}& \colhead{}& \multicolumn{2}{c}{2004 May 13.6}& & \multicolumn{2}{c}{2007 Mar. 26.2}& & \multicolumn{2}{c}{2008 Feb. 25.4} \\[0.5ex]
\cline{4-5} \cline{7-8} \cline{10-11}\\[-1.8ex]
\colhead{Ion}& \colhead{(\micron)}& \colhead{Module}& Flux\tablenotemark{a}& FWHM\tablenotemark{b}& & Flux\tablenotemark{a}& FWHM\tablenotemark{b}& & Flux\tablenotemark{a}& FWHM\tablenotemark{b} 
}

\startdata

[O IV]& 25.91& LH& $70.1 \pm 0.3$& $2840 \pm 10$& & $17.9 \pm 0.3$& $2530 \pm 30$& & $11.3 \pm 0.1$& $2510 \pm 20$\\

[Ne II]& 12.81& SL,SH& $16.6 \pm 0.3$& $2140 \pm 20$& & $5.6 \pm 0.3$& $2160 \pm 70$& & $4.2 \pm 0.1$& $2200 \pm 30$ \\

[Ne III]& 15.56& SH& $61.2 \pm 0.6$& $2360 \pm 20$& & $17.7 \pm 0.4$& $2370 \pm 30$& & $11.7 \pm 0.1$& $2290 \pm 12$ \\

[Ne III]& 36.01& LH& $4.1 \pm 1.0$& $2000 \pm 400$& & -- & -- & & $1.0 \pm 0.3$& $1700 \pm 500$ \\

[Ne V]& 14.32& SH& $9.7 \pm 0.3$& $2460 \pm 60$& & -- & -- & & -- & -- \\

[Ne V]& 24.30& LH& $6.9 \pm 0.2$& $2450 \pm 40$& & -- & -- & & -- & -- \\

[S IV]& 10.51& SL,SH\tablenotemark{c}& $0.69 \pm 0.20$& $4338 \pm 935$& & -- & -- & & -- & -- \\

[Ar III]& 8.99& SL& $0.34 \pm 0.30$& $4733 \pm 3050$& & -- & -- & & -- & -- \\[-2.5ex]

\enddata

\tablenotetext{a}{Fluxes provided in units of $10^{-13}$ erg s$^{-1}$ cm$^{-2}$}
\tablenotetext{b}{FWHM velocity widths in km s$^{-1}$, uncorrected for the instrument resolution of $\sim 500$ km s$^{-1}$.}
\tablenotetext{c}{When both high- and low-resolution data were available, we preferentially selected the high-res data for measurement of line parameters.}

\end{deluxetable*}


Emission from [\ion{Ne}{2}] 12.81 \micron\ was relatively strong with a saddle shape that persisted  throughout our observations. The [\ion{Ne}{3}] 15.56 \micron\ emission was nearly four times the intensity of [\ion{Ne}{2}]. The saddle-shaped structure in this feature was less well defined, but was similar to the [\ion{Ne}{2}] line in that it was dominated by a blue-ward peak. The [\ion{Ne}{5}] lines at 14.32 and 24.30 \micron\ were apparent only during the first epoch of observations, after which they were no longer detected. Their line profiles were rounded, distinctly different from the [\ion{Ne}{2}] and [\ion{Ne}{3}] lines. The [\ion{O}{4}] 25.91 \micron\ line was the strongest in the spectra. The profile shape was rounded like the [\ion{Ne}{5}] lines. The IRS high-res (R $\sim 600$) profile of the [\ion{S}{4}] 10.51 \micron\ line was too noisy to be accurately characterized. 

\subsubsection{V382Vel - Nebular Environment}
\label{sec:V382Vel_Line_Struct}

The presence of [\ion{Ne}{2}] and [\ion{Ne}{3}] in the same spectra as [\ion{Ne}{5}] is indicative of ionization gradients in the ejecta, which suggest strong density inhomogeneities. This is consistent with evidence from earlier in the eruption suggesting that the fragmentation of the ejecta was fixed early in the outburst \citep{Shore03}. There is substantial evidence from both direct imaging of resolved nova shells \citep[see][and references therein]{OBrienBode08} and photoionization modeling \citep[cf.][]{Helton10a,Vanlandingham05} that the ejecta of CNe often exhibit substantial density inhomogeneities. It is therefore unsurprising that V382 Vel would exhibit similar structure and the fact that the line profiles remained unchanged over the 3.8 year coverage of our observations reveals that the kinematic structure of the ejecta components was stable.  

The difference in the line shapes of the low ionization lines in comparison to the high ionization lines implies that they arise in physically distinct regions of the ejecta. Figure \ref{fig:FWHMEvolve} shows that the lines with lower I.P. have narrower FWHM velocities increasing up to that of [\ion{O}{4}]. The higher I.P. [\ion{Ne}{5}] line was slightly narrower than [\ion{O}{4}], but the difference is probably not significant. One interpretation is that the low I.P. [\ion{Ne}{2}] emission arises in the low velocity, dense component and the [\ion{Ne}{5}] emission comes from higher velocity, more diffuse material. Indeed, the [\ion{Ne}{3}] profile exhibits characteristics intermediate between those of the [\ion{Ne}{2}] and [\ion{Ne}{5}] transitions, as would be expected for emission arising from the transition area between the two components. Based upon the profile shapes, we hypothesize that the [\ion{O}{4}] emission is arising from the same region in the ejecta as the [\ion{Ne}{5}] emission. 

Figure \ref{fig:FluxEvolve} reveals that the line fluxes of [\ion{Ne}{2}], [\ion{Ne}{3}], and [\ion{O}{4}] declined with a shallow slope, $\alpha = 2.5 \pm 0.1, 2.9 \pm 0.1, $ and $3.2 \pm 0.1$ respectively. These slopes are more shallow than would be expected for an expanding thin shell, again supporting the hypothesis that the ejecta have a clumpy morphology. In addition, if the shallowness of the decline is indicative of the degree of clumpiness or fragmentation of the emitting material, then one would expect that [\ion{Ne}{2}] traces the most highly fragmented material, followed by [\ion{Ne}{3}], and then [\ion{O}{4}]. We do not have enough information to characterize the [\ion{Ne}{5}] emission since it was unobserved in our final two epochs of observations.

Taken together, the above arguments about the emission line structure may lead to a broad understanding of the ejecta structure without recourse to direct imaging of the nova shell. The profile shapes, widths, and fluxes all point to a shell model of the ejecta in which the most highly ionized material is nearly spherically symmetric and lies exterior to the slower, slightly more dense and clumpy material that gives rise to the low I.P. emission lines. This latter interior component seems to be more geometrically thin and to be more asymmetric than the diffuse, highly ionized material. The [\ion{Ne}{3}] emission traces the transition between these two ejecta components. This interpretation is consistent with those presented by others, including \citet{DellaValle02}, who suggested that the ejecta were distributed in a discrete shell, and \citet{Shore03}, who noted that the high velocity material appeared to be more spherically symmetric than the low velocity material.

\subsubsection{V382 Vel - Adopted Parameters}
\label{sec:V382_AV}

The distance to V382 Vel has been well constrained through use of the MMRD relationship and various qualitative arguments. These estimates are summarized in Table \ref{tab:distances}. For the following analysis, we use a distance of 2.3 kpc.

The mass of hydrogen in the ejecta has been estimated using a variety of methods. \citet{Shore03} estimated an ejected mass of M$_{ej}=(4-5) \times 10^{-4}$ M$_{\odot}$ based upon photoionization models of their UV spectra. In comparison, \citet{DellaValle02} derived a mass of $6.5 \times 10^{-6}$ M$_{\odot}$ using the luminosity of H$\alpha$ assuming a distance of 1.7 kpc. Scaling this estimate to a distance of 2.3 kpc results in a mass of approaching $10^{-5}$ M$_{\odot}$. We adopt a mass of $4 \times 10^{-4}$ M$_{\odot}$ for the ejecta, which results in a lower limit on the calculated abundances. 

In order to adequately constrain the electron density in the shell, we must determine the appropriate expansion velocity of the emitting region.  The P-Cyg absorption components observed early in the development exhibited a range of terminal velocities, from 2300 to 5000 km s$^{-1}$ \citep{DellaValle02}. These high expansion velocities were corroberated by early emission line measurements having HWZI $\gtsimeq 3500$ km s$^{-1}$. We measure the minimum expansion velocity of 1200 km s$^{-1}$ from the HWHM of the late-time \textit{Spitzer} spectra and we adopt 3500 km s$^{-1}$ as a conservative estimate on the maximum expansion velocity. 

In our abundance analysis (\S\ref{sec:V382_Ab}), we present two model cases for the ejecta. The first assumes a spherical shell defined by a minimum expansion velocity of $v_{min}=1200$ km s$^{-1}$ and a maximum of $v_{max}=3500$ km s$^{-1}$, resulting in n$_{e} \sim 700$ cm$^{-3}$. The second case consists of a spherical shell with an expansion velocity at the leading edge of $v_{max}=3500$ km s$^{-1}$ and a thickness of $\Delta r = 0.1r_{out}$. This model results in an electron density of n$_{e} \sim 2500$ cm$^{-3}$.

\begin{figure}
\epsscale{1.2}
\plotone{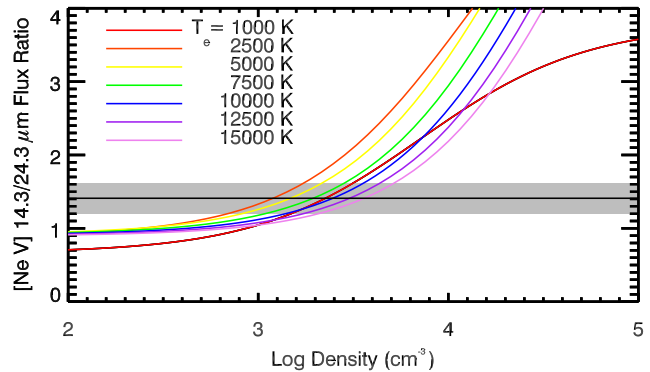}
\caption{This plot shows the temperature and density dependence of the [\ion{Ne}{5}] 14.32/24.30 $\micron$ line ratio. The plots of line ratio vs.\ $n_{e}$ are for electron temperatures ranging from 1000 to 15000 K. The horizontal line is the measured flux ratio for V382 Vel with the error on the ratio indicated by the grey shaded region. \label{fig:NeVTemDen}}
\end{figure}

In order to estimate the electron temperature of the ejecta, in Figure \ref{fig:NeVTemDen} we plot the theoretical [\ion{Ne}{5}] 14.32/24.30 \micron\ ratio as a function of $n_{e}$ for $T_{e}$ = 1000, 3000, 5000, and 10000 K. The measured value for V382 Vel is indicated by the solid horizontal line. The grey shaded area accounts for the error in the line ratio arising from the errors on the line fits. From the above geometrical arguments, we expect the electron density to be $2.85 \leq \mathrm{log(n}_{e}\mathrm{)} \leq 3.40$. At the low density limit, the [\ion{Ne}{5}] ratio is consistent with T$_{e} \sim 3000$ K, while at the high density extreme, the temperature is double valued at $\sim 1000$ K or $\sim 10000$ K. The errors on the [\ion{Ne}{3}] 15.56/36.01 \micron\ ratio are much higher than on the [\ion{Ne}{5}] ratio and consequently provide much weaker constraints on the temperature. They are, however, more consistent with a temperature less than 5000 -- 10000 K. For subsequent analysis, we bracket our abundance estimates with temperatures of 1000 and 5000 K. 

The adopted parameters are summarized in Table \ref{tab:parameters}.

\subsubsection{V382 Vel - Abundances}
\label{sec:V382_Ab}

In Table \ref{tab:v382_ab} we present the calculated elemental abundances by number relative to solar for the observed neon, oxygen, argon and sulfur lines for four different combinations of electron density and temperature. Again, the calculated abundance solutions have a stronger dependence on $n_{e}$ than $T_{e}$. The derived Ne abundance, calculated by combining the implied abundances from each of the observed emission lines, ranges from 18 times solar at T$_{e} = 5000$ K and n$_{e} \sim 2500$ cm$^{-3}$ (20 times solar at 1000 K) to 61 times solar at T $_{e} = 5000$ K and n$_{e} \sim 700$ cm$^{-3}$ (70 times solar at 1000 K). Similarly, the lower limits to the oxygen abundance ranged from 1.7 to 3.9 times solar by number relative to hydrogen, depending upon the assumed density. The abundances of argon and sulfur were low, with abundances by number relative to hydrogen of 0.1 - 0.5 and 0.1 - 0.3 times solar, respectively. Since these lower limits to the abundances are below the solar values, the results for these species remain inconclusive.


\begin{deluxetable*}{lccccccc}
\tabletypesize{\scriptsize}
\setlength{\tabcolsep}{0.025in}

\tablewidth{0pt}
\tablecaption{Abundances by Number Relative to Solar - V382 Vel \label{tab:v382_ab}}

\tablehead{
\colhead{}& \colhead{}& \colhead{}& \multicolumn{5}{c}{Derived Abundances\tablenotemark{a}}\\[0.5ex]
\cline{4-8}\\[-1.8ex]
\colhead{Species}& \colhead{Wavelength}& \colhead{$n_{e}$ ($cm^{-3}$):}& \multicolumn{2}{c}{700}& \colhead{}& \multicolumn{2}{c}{2500}\\
\cline{4-5} \cline{7-8}\\[-1.5ex]
\colhead{}& \colhead{(\micron)}& \colhead{$T_{e}$ (K):}&
\colhead{ \ 1000 \ }& \colhead{ \ 5000 \ }& \colhead{}& \colhead{ \ 1000 \ }& \colhead{ \ 5000 \ }
}

\startdata

[Ne II]& 12.81& & -2.78& -2.83& & -3.33& -3.38\\

[Ne III]& 15.56& & -2.38& -2.44& & -2.92& -2.99\\

[Ne V]& 14.32& & -4.34& -4.28& & -4.77& -4.71\\

& 24.30& & -4.15& -4.02& & -4.51& -4.40\\[0.5ex]

\hline\\[-1.5ex]

\textbf{Total Neon}& & & -2.23 (70.0)& -2.28 (61.3)& & -2.76 (20.2)& -2.82 (17.8)\\[0.5ex]

\hline\\[-1.5ex] 

[O IV]& 25.91& & -2.75 (3.6)& -2.72 (3.9)& & -3.09 (1.7)& -3.04 (1.9)\\

[Ar III]& 8.99& & -5.87 (0.5)& -6.07 (0.3)& & -6.40 (0.2)& -6.60 (0.1)\\

[S IV]& 10.51& & -5.43 (0.3)& -5.70 (0.2)& & -5.93 (0.1)& -6.18 (0.1)\\[-2.5ex]

\enddata

\tablenotetext{a}{The number given is the log abundance by number relative to hydrogen. The number in parentheses is the abundance by number relative to hydrogen, relative to solar assuming the solar values of \citet{Asplund09}.}
\tablenotetext{b}{When both high and low resolution data are available, we preferentially select the high res data for the estimation of the total abundances.}

\end{deluxetable*}


The lower limits to the abundances derived here are consistent with estimates made by other authors, provided that the system is described by our higher density model. \citet{AD03} estimated lower limits to the abundances of O, Ne, S, and Ar using primarily optical data from 8 epochs spanning roughly 70 to 700 days after outburst using the IRAF task {\scriptsize NEBULAR.IONIC}. They found that Ne was $\geq 12$ times solar by number. Their lower limit for oxygen, however, was only 0.3 times solar while they estimated abundances of S and Ar of $\geq 4.9$ and $\geq 0.8$ times solar, respectively. \citet{Shore03} used the Cloudy photoionization code \citep{Fer98} to model the UV emission lines observed 110 days post-outburst and found that O and Ne were significantly overabundant. They derived minimum oxygen and Ne abundances by number relative to solar of $3.4 \pm 0.3$ and $17 \pm 3$ respectively. Together these three sets of abundance estimates support the conclusion of \citet{AD03}, that V382 Vel is likely an ``intermediate neon nova'' with outburst properties mimicking classical ONe novae such as QU Vul and V1974 Cyg, but with a somewhat lower Ne abundance.

\subsection{V1494 Aql (Nova Aquila 1999 No. 2)}
\label{sec:V1494Aql}

V1494 Aql was discovered by A. Pereira on 1999 December 1.785 UT at a pre-maximum visual magnitude of m$_{\mathrm{V}} \sim 6.0$ \citep{Pereira99}. \citet{KissThomson00} reported that the nova reached maximum at m$_{\mathrm{V}} =  4.0$ mag roughly one day later on December 3.4 UT (t$_{0}$; JD 2451515.9). They found that the light curve declined rapidly with t$_{2} = 6.6 \pm 0.5$ days and t$_{3} = 16.0 \pm 0.5$ days, which situates the nova in the ``very fast'' speed class of \citet{Gaposchkin57}. The light curve was initially smooth, but during the transition stage it exhibited strong oscillations of $\sim 1.2$ mag peak-to-trough in V-band with a period of approximately $16.5 \pm 0.1$ days \citep{IijimaEsenoglu03}. By virtue of these oscillations, \citet{Strope10} categorized V1494 Aql as a rare O-type system. Even amongst this small group, which includes only about 4\% of CNe, V1494 Aql stands out due to the irregularity in the period and shape of its oscillations. In this respect, its light curve is most similar to that of V888 Cen. There was no evidence in the light curve for dust production in the ejecta of V1494 Aql.

Optical spectroscopy during the early stages of development revealed a strong hydrogen recombination spectrum along with emission from \ion{O}{1} $\lambda$7773 \AA\ and \ion{Fe}{2} \citep{Fujii99}, the latter resulting in the classification of V1494 Aql as an ``Fe II''-type nova. Strong P-Cygni absorption components were observed on the Balmer series at -1850 km s$^{-1}$ \citep{Moro99} and weaker absorption systems on the \ion{O}{1} and \ion{Fe}{2} lines at velocities of -1020 km s$^{-1}$ \citep{Ayani99}. \citet{IijimaEsenoglu03} calculated a mean expansion velocity from the absorption components of -2510 km s$^{-1}$. Within a few days of maximum, the P-Cyg absorption components disappeared, leaving behind rounded emission lines characteristic of an optically thin wind \citep{Anupama01,Kamath05}. These emission lines rapidly developed a double peaked, saddle-shaped structure that was attributed to an equatorial-ring/polar-cap morphology of the ejecta \citep{KissThomson00,Anupama01} based upon the synthetic models of \citet{GillOBrien99}. Spectropolarimetric observations very early in the evolution supported the conclusion that the ejecta were asymmetric and that the asymmetry was likely established during the pre-maximum rise \citep{IijimaEsenoglu03}. The fine scale structure in the saddle-shaped profiles of [\ion{O}{3}], [\ion{N}{2}], and H$\beta$ was nearly identical, indicating that these lines formed from the same location in the ejecta and that information gathered from their profiles can likely be used to characterize the emitting region quite well. 

The system evolved to the coronal stage by $\sim 150$ days after outburst with the appearance of [\ion{Fe}{10}] at a level greater than [\ion{Fe}{7}] in the optical \citep{Kamath05}. Interestingly, at the onset of the coronal stage, round-topped line profiles were observed in \ion{N}{2} and \ion{N}{3} in striking contrast to the saddle-shaped profiles that persisted in the nebular lines \citep{IijimaEsenoglu03}. As the coronal spectrum matured, the emission lines from \ion{He}{2} and the highly ionized species of iron, such as [\ion{Fe}{10}] and [\ion{Fe}{11}], also transitioned to smooth, rounded profiles. \citet{IijimaEsenoglu03} suggested this might have been due to the presence of an additional optically thin wind with a spherically symmetric, uniform distribution. The optical spectra during this stage were dominated by [\ion{O}{3}] and numerous forbidden iron lines. Although there was a weak signature of emission from sulfur, there are no reports in the literature of neon emission. 

The underlying radiation during the coronal stage was extremely energetic. \citet{Mazuk00} reported emission from [\ion{S}{9}] 1.2523 \micron\ (I.P. = 329 eV) on day 226, while \citet{Rudy01} observed [\ion{S}{11}] 1.9196 \micron ~(I.P. = 447 eV) on day 580. X-ray observations indicated that during this period the underlying source exhibited characteristics similar to super-soft sources \citep{Drake03,Rohrbach09} as would be expected for sources exhibiting [\ion{Fe}{10}] emission \citep{Schwarz11}.

\subsubsection{V1494 Aql - Optical Data}
\label{sec:V1494Aql_Opt}

\begin{figure*}
\epsscale{1.1}
\plotone{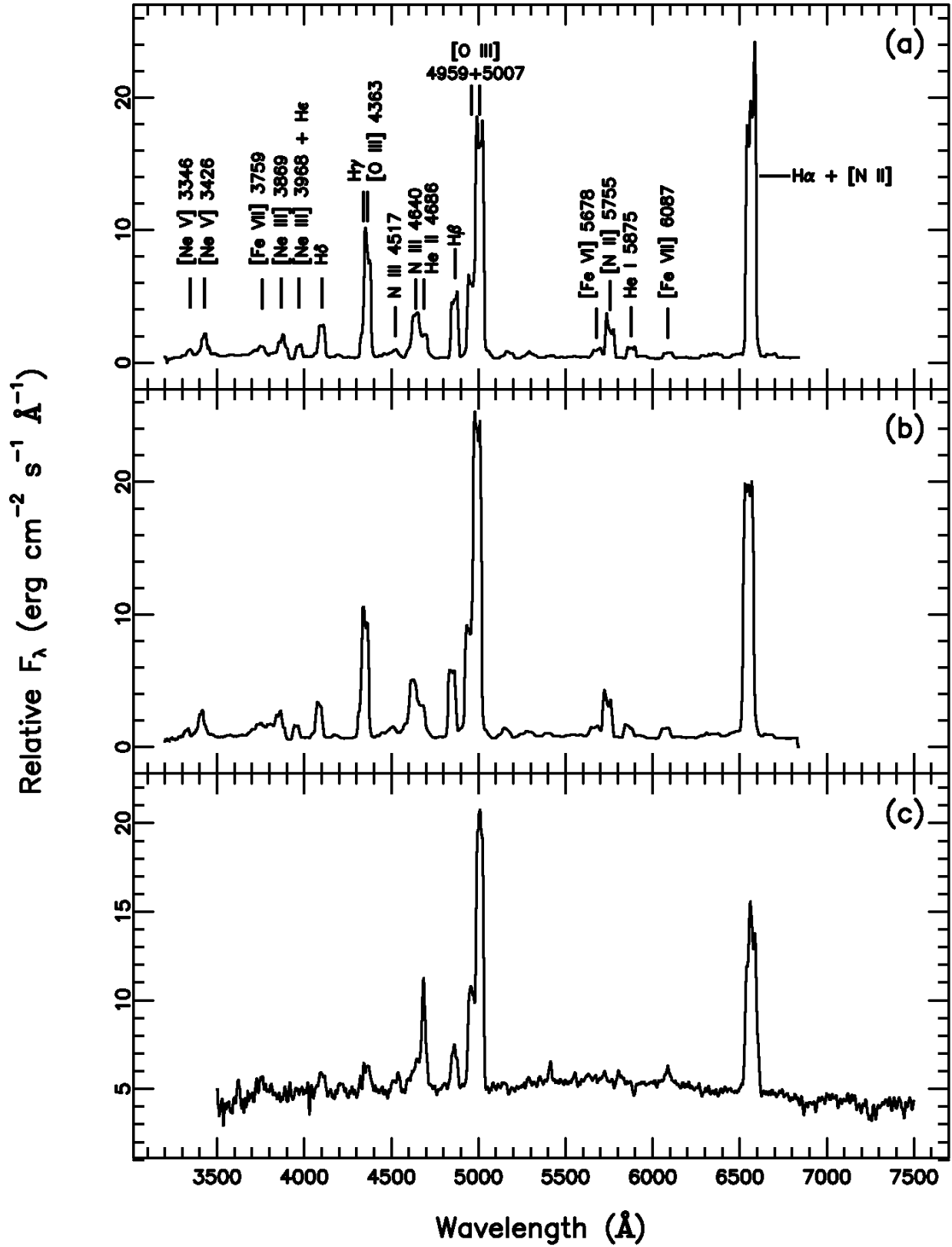}
\caption{V1494 Aql:  Optical spectral evolution of V1494 Aql. The spectra have not been corrected for interstellar extinction and reddening. (a) Spectrum obtained on 2000 April 25.5 UT (Day 143). (b) Spectrum obtained on 2000 June 4.5 (Day 185).  This spectrum is very similar to that obtained on 2000 April 4.5.  (c)  Spectrum obtained on 2004 June 25 (Day 1666).  \label{fig:V1494Aql_Opt}}         
\end{figure*}

The optical spectra of V1494 Aql we obtained using the Hiltner 2.4-m and the Bok 2.3-m telescopes are shown in Figure \ref{fig:V1494Aql_Opt}. The first two epochs of data, obtained 143 and 185 days after maximum, reveal a system well into the nebular stage of development, dominated by hydrogen recombination lines and strong lines of [\ion{O}{3}] $\lambda\lambda$4959, 5007 and $\lambda$4363 \AA. Also present are broad and strong permitted emission lines of \ion{He}{1}, \ion{He}{2}, and \ion{N}{3} as well as forbidden emission lines of [\ion{Fe}{7}], [\ion{N}{2}], [\ion{Fe}{6}], [\ion{O}{3}], [\ion{Ne}{3}], and [\ion{Ne}{5}]. The lines have broad, flat topped or saddle-shaped profiles consistent with those seen earlier in the outburst \citep[e.g.,][]{KissThomson00,Anupama01,IijimaEsenoglu03}. At this phase the electron densities are relatively high.

The observations of [\ion{Ne}{3}] and [\ion{Ne}{5}] are the first reported detections of neon in V1494 Aql. We suspect that this is because observations by other authors did not extend to short enough wavelengths to see [\ion{Ne}{3}] or [\ion{Ne}{5}] and that any emission of [\ion{Ne}{4}] $\lambda\lambda$4714, 4725 \AA\ that may have been present was either missed because the line was short lived and the data possessed inadequate temporal coverage or the line was too heavily blended with other emission features to be clearly identified. This emphasizes the need for early spectra extending to at least 3800 \AA, obtained with a high temporal cadence.

The only possible detection of neon to be found in the literature is in the spectrum presented by \citet{IijimaEsenoglu03} from 2000 Feb. 6 (Day 65; their Figure 10), which revealed partial coverage of a broad, flat-topped emission line that is probably due to [\ion{Ne}{3}] $\lambda$3968 \AA\ blended with H$\epsilon$. Based upon the partial line profile and assuming that the H$\epsilon$ contribution is consistent with Case B recombination, we estimate that the flux from [\ion{Ne}{3}] was probably no more than 20\% of H$\beta$. This is lower than might be expected for a typical ONe nova during the nebular stage.

By the third epoch (Fig. \ref{fig:V1494Aql_Opt}, panel c), 1666 days after maximum, many of the emission lines had faded substantially. Some of the iron lines observed earlier may still be present at very low levels, but if so, their line profiles had narrowed and were no longer flat-topped. The \ion{He}{1} lines were no longer present, while the \ion{He}{2} feature at 4686 \AA\ had strengthened relative to H$\beta$ and taken on a very narrow, rounded profile. The [\ion{Ne}{3}] doublet was no longer apparent. The strongest feature in the spectrum is the [\ion{O}{3}] line at 5007 \AA. The [\ion{O}{3}] profiles had transitioned from the castellated shapes observed at the early epochs to rounded shapes, similar to that seen in the hydrogen lines. A weak feature near 3727 \AA\ might correspond to [\ion{O}{2}] consistent with the low densities inferred from the contemporaneous Spitzer spectra (see \S\ref{sec:V1494Aql_IR}). 

\subsubsection{V1494 Aql - \textit{Spitzer} Data}
\label{sec:V1494Aql_IR}

\begin{figure*}
\epsscale{1.2}
\plotone{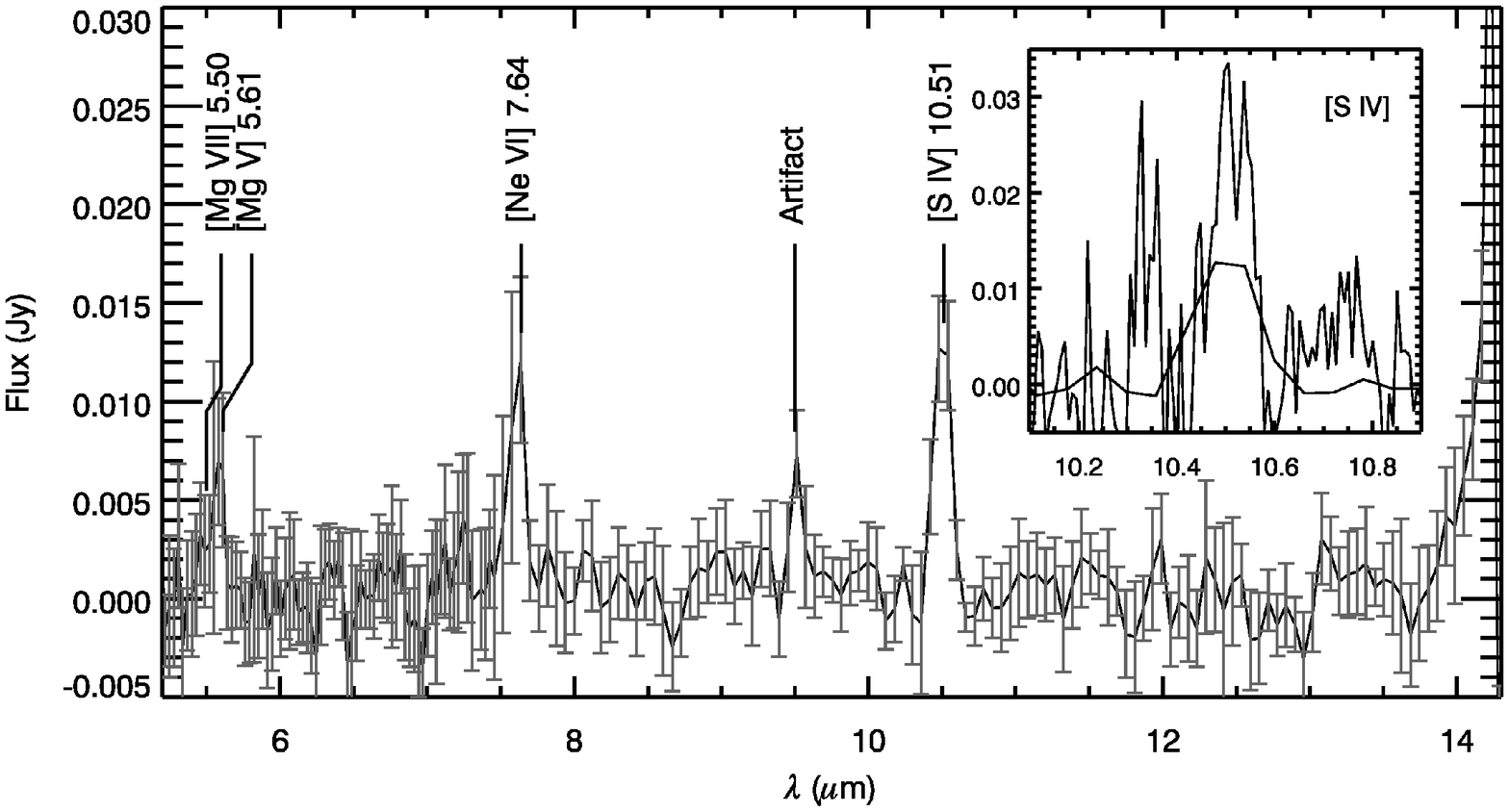}
\caption{V1494 Aql: \textit{Spitzer} SL spectra of V1494 Aql obtained on 2004 April 14. Prominent emission lines are labeled. \textit{Inset}: Comparison of the region around [\ion{S}{4}] 10.51 \micron\ in the SL module to the SH module. Since the SH data were not background corrected, they were offset to match the continuum level of the SL data. \label{fig:V1494Aql_Low}}
\end{figure*}

\begin{figure}
\epsscale{1.1}
\plotone{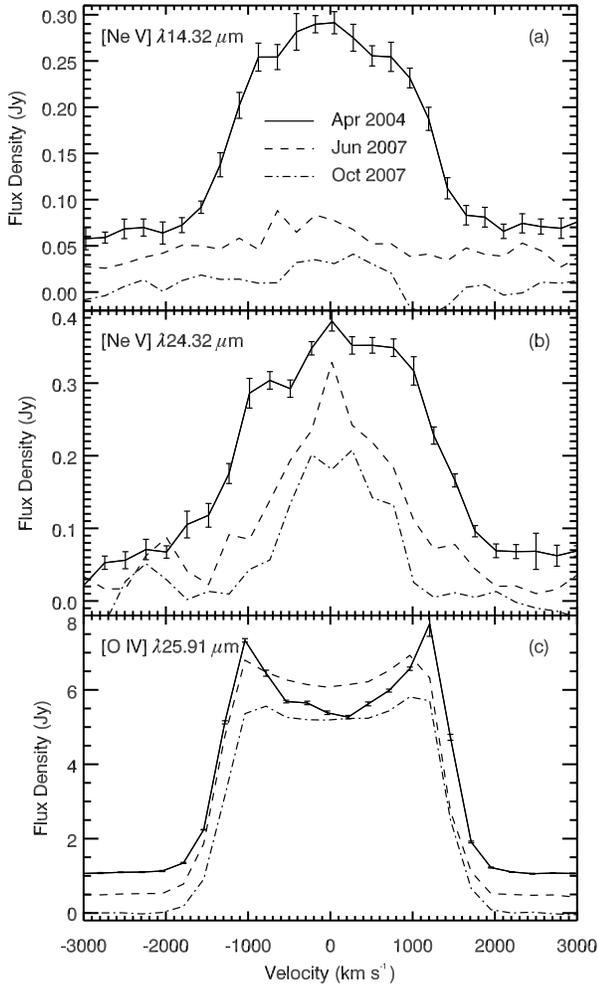}
\caption{Plotted above are line profiles for the dominant emission lines in V1494 Aql from three epochs of observations. Representative errors are plotted on the profiles for 14 April 2004. Arbitrary bias levels have been applied to the April 2004 and June 2007 data for clarity. \textit{Panel (a)}: This panel shows the profiles of [\ion{Ne}{5}] 14.32 \micron. \textit{Panel (b)}: This panel shows the region around [\ion{Ne}{5}] 24.32 \micron. The June 2007 and October 2007 spectra have been scaled by a factor of 3.0. \textit{Panel (c)}: This panel presents the profiles of [\ion{O}{4}] 25.91 \micron. The June 2007 and October 2007 spectra have been scaled by a factor of 3.0. \label{fig:V1494Aql_MIR}}
\end{figure}

We obtained mid-IR spectra during three epochs as part of our \textit{Spitzer} CN monitoring campaign. These data were obtained roughly 4.4, 7.5, and 7.9 years after maximum. The spectra are shown in Figures \ref{fig:V1494Aql_Low} and \ref{fig:V1494Aql_MIR}. The first observation was obtained only 72 days before the third epoch of optical data.  The \textit{Spitzer} spectra from the first visit are dominated by emission lines of [\ion{O}{4}] at 25.91 \micron, [\ion{Ne}{5}] 14.32 and 24.30 \micron, and [\ion{S}{4}] 10.51 \micron. There is also weak emission from [\ion{Ne}{6}] 7.61 \micron, [\ion{Mg}{5}] 5.61 \micron, and possibly [\ion{Mg}{7}] 5.50 \micron. A summary of the line measurements are provided in Table \ref{tab:V1494Aql_Lines}.

The presence of strong [\ion{Ne}{5}] lines in the IR in conjunction with the disappearance of the [\ion{Ne}{3}] lines in the optical suggests that the overall level of ionization of the ejecta had increased dramatically by Day 1666. This is supported by the increasing relative strength observed in the optical \ion{He}{2} line.

\begin{deluxetable*}{lcccccccccc}
\tabletypesize{\scriptsize}
\setlength{\tabcolsep}{0.05in}

\tablewidth{0pt}
\tablecaption{V1494 Aql - Line Measurements \label{tab:V1494Aql_Lines}}

\tablehead{
\colhead{}& \colhead{$\lambda_{o}$}& \colhead{}& \multicolumn{2}{c}{2004 Apr. 14.3}& & \multicolumn{2}{c}{2007 Jun. 8.2}& & \multicolumn{2}{c}{2007 Oct 12.3} \\[0.5ex]
\cline{4-5} \cline{7-8} \cline{10-11}\\[-1.8ex]
\colhead{Ion}& \colhead{(\micron)}& \colhead{Module}& Flux\tablenotemark{a}& FWHM\tablenotemark{b}& & Flux\tablenotemark{a}& FWHM\tablenotemark{b}& & Flux\tablenotemark{a}& FWHM\tablenotemark{b} 
}

\startdata

[O IV]& 25.91& LH& $57.3 \pm 0.3$& $2872 \pm 6$& & $20.8 \pm 0.2$& $2760 \pm 16$& & $18.7 \pm 0.2$& $2730 \pm 19$\\

[Ne V]& 14.32& SH& $3.6 \pm 0.2$& $2100 \pm 100$& & $0.6 \pm 0.3$& $1800 \pm 700$& & -- & -- \\

[Ne V]& 24.30& LH& $3.0 \pm 0.2$& $2100 \pm 100$& & $0.6 \pm 0.1$& $1400 \pm 200$& & $0.4 \pm 0.1$& $1300 \pm 200$ \\

[Ne VI]& 7.64& SL& $0.5 \pm 0.6$& $2880 \pm 2955$& & -- & -- & & --& -- \\

[Mg V]& 5.61& SL& $0.6 \pm 0.5$& $5086 \pm 1259$& & -- & -- & & -- & -- \\

[Mg VII]& 5.50& SL& $0.3 \pm 0.4$& $5184 \pm 1283$& & -- & -- & & -- & -- \\

[S IV]& 10.51& SL,SH\tablenotemark{c}& $0.6 \pm 0.4$& $2067 \pm 937$& & -- & -- & & -- & -- \\[-2.5ex]

\enddata

\tablenotetext{a}{Fluxes provided in units of $10^{-13}$ erg s$^{-1}$ cm$^{-2}$}
\tablenotetext{b}{FWHM velocity widths in km s$^{-1}$, uncorrected for the instrument resolution of $\sim 500$ km s$^{-1}$.}
\tablenotetext{c}{When both high- and low-resolution data were available, we preferentially selected the high-res data for measurement of line parameters.}

\end{deluxetable*}


\subsubsection{V1494Aql - Nebular Environment}
\label{sec:V1494Aql_Line_Struct}

Clues to the origins of the optical emission lines can be derived from their profiles. The oxygen line at 25.91 \micron\ has the same saddle-shaped structure present in most of the optical and near-IR emission lines reported during the first few hundreds of days after outburst, which is claimed to be characteristic of an equatorial-ring/polar-cap morphology \citep{KissThomson00, Anupama01}. The neon lines, on the other hand, have smooth bullet-shaped profiles, similar to the profiles of the \ion{He}{2} and highly ionized iron lines observed during the coronal stage, which suggests that they may arise in the optically thin, geometrically thick shell or wind. 

\begin{figure}
\epsscale{1.2}
\plotone{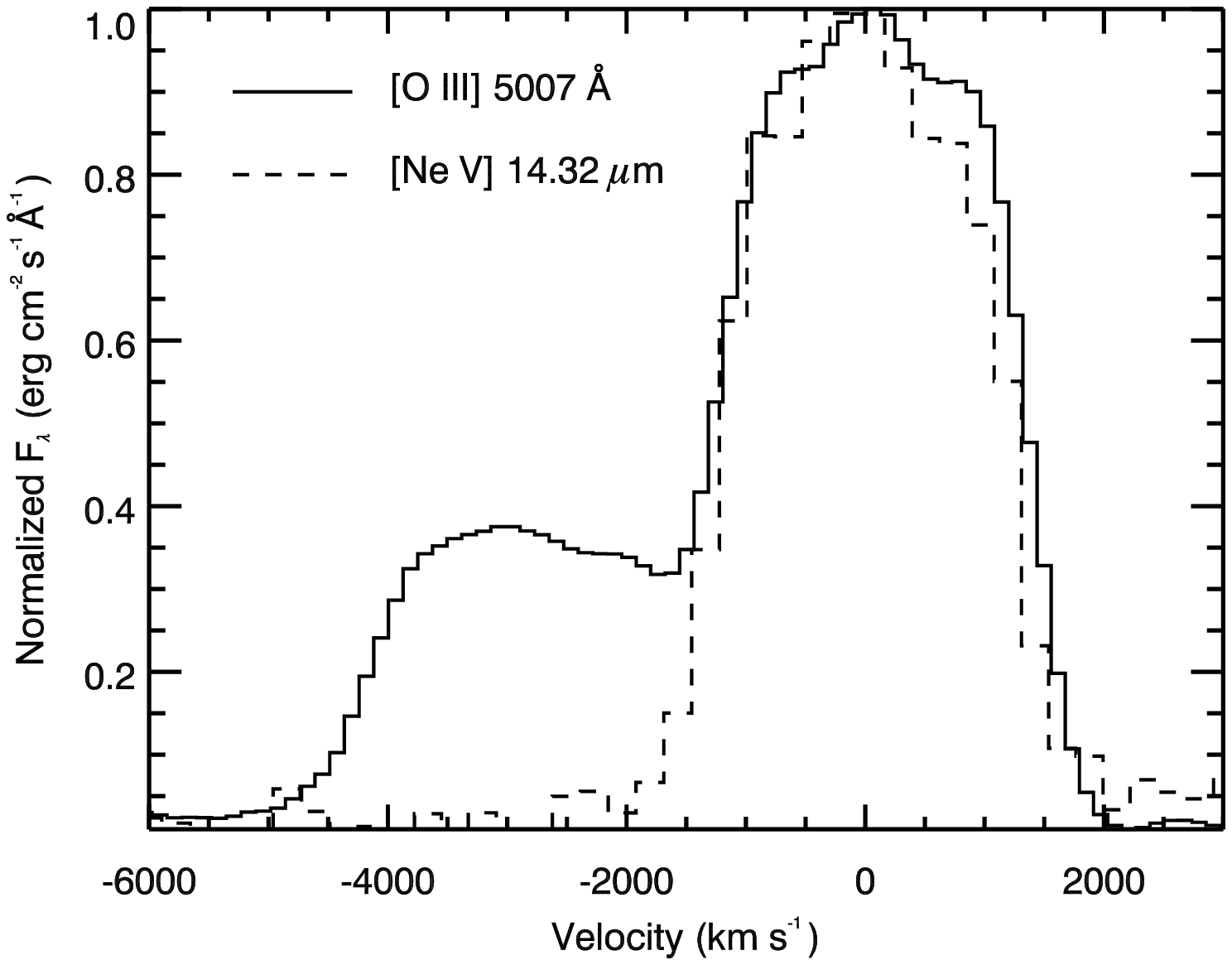}
\caption{V1494 Aql: Enlargement of the spectral region near H$\beta$ from the spectrum shown in Figure \ref{fig:V1494Aql_Opt}.  The [\ion{O}{3}] 5007 \AA\ line profile exhibits nearly the identical saddle-shaped structure as the [\ion{Ne}{5}] 24.32 \micron\ line profile (Figure \ref{fig:V1494Aql_MIR}). \label{fig:V1494Aql_ProfCompare}}
\end{figure}

At late times, the conditions in the ejecta can be inferred by comparison of the optical and IR emission lines. The structure of the [\ion{O}{3}] is the same as that of the [\ion{Ne}{5}] lines in the IR. This is displayed in Figure \ref{fig:V1494Aql_ProfCompare}. Therefore, it is probably safe to assume that the regions giving rise to these two species have the same bulk geometry, probably a spherical shell. The behavior of the [\ion{Ne}{5}] line flux decline, $\propto t^{-4}$ (Figure \ref{fig:FluxEvolve}), is consistent with an optically thin shell.

\begin{figure}
\epsscale{1.2}
\plotone{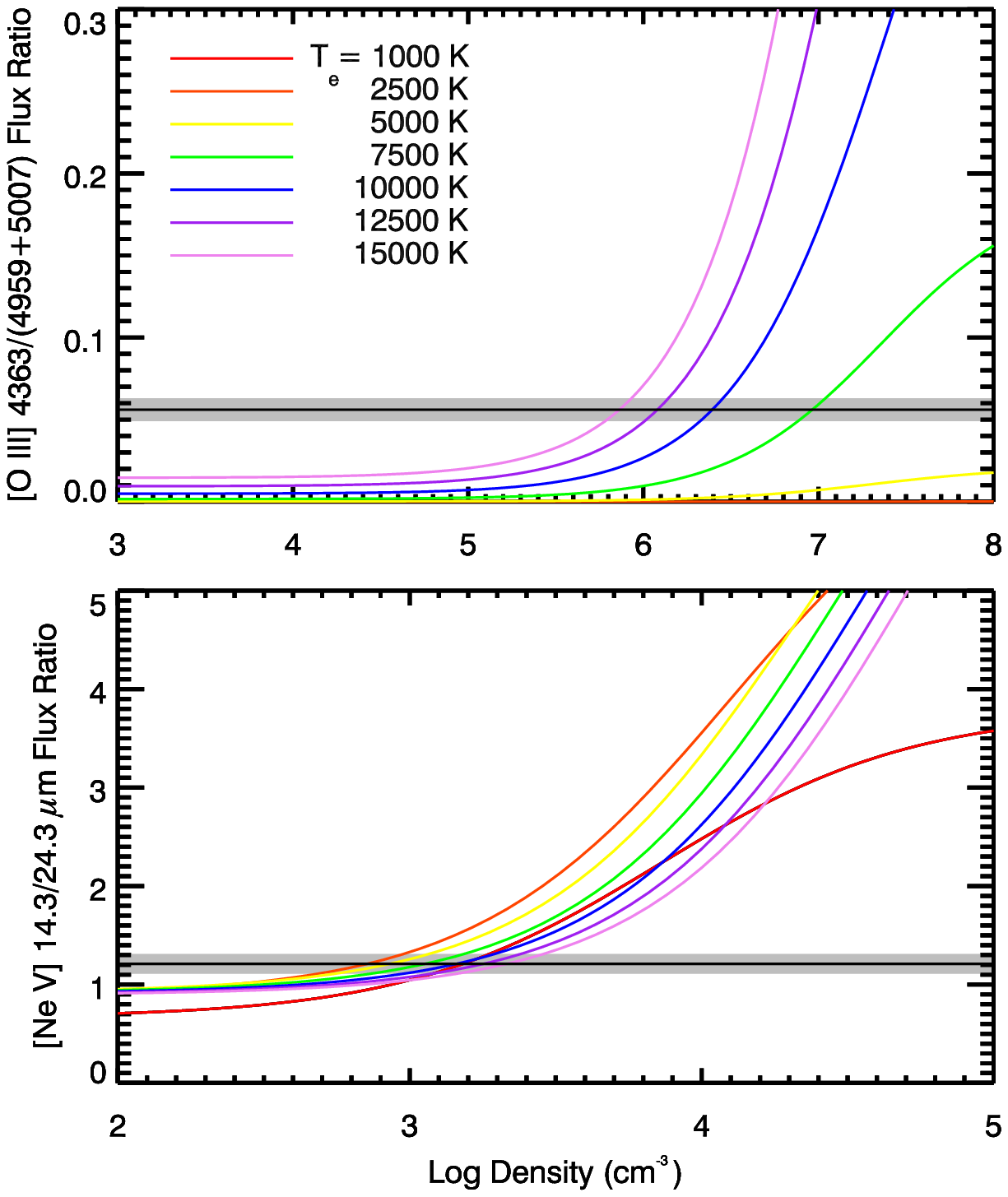}
\caption{This plot shows the temperature and density dependence of the [\ion{O}{3}] 4363/(4959 + 5007) \AA\ line ratio (\textit{top}) and the [\ion{Ne}{5}] 14.32/24.30 $\micron$ line ratio (\textit{bottom}). The predicted line ratio vs.\ $n_{e}$ are color coded for electron temperatures ranging from 1000 to 15000 K. The horizontal lines are the measured flux ratios and the grey shaded area indicates the formal error on the ratio.  \label{fig:V1494Aql_TemDen}}
\end{figure}

If the [\ion{Ne}{5}] and [\ion{O}{3}] emitting regions have the same global geometry, then we would expect that they would also have nearly the same temperatures. For our analysis, we assume that the temperature is no higher than $1.5 \times 10^{4}$ K, a reasonable upper limit for photoionization. In Figure \ref{fig:V1494Aql_TemDen}, we plot the predicted ratios of the optical [\ion{O}{3}] lines at 4363, 4959, and 5007 \AA\ and the IR [\ion{Ne}{5}] lines at 14.32 and 24.30 \micron\ as a function of density for several values of $T_{e}$. The error in the [\ion{O}{3}] ratio is dominated by errors incurred from deblending [\ion{O}{3}] $\lambda$4363 from H$\gamma$. For the chosen range of nebular temperatures, the derived densities based on the [\ion{O}{3}] line ratios are in the range $6.1 \lesssim$ log($n_{e}$) $\lesssim 7.1$. In the case of the optical Ne lines, the required density of the [\ion{Ne}{5}] emitting region is of order $2.7 \lesssim$ log($n_{e}$) $\lesssim 3.3$. These latter density estimates are generally consistent with the densities estimated from geometrical arguments based upon the ejecta dynamics and ejecta masses reported in the literature (see \S\ref{sec:V1494_AV} below). However, densities derived from the [\ion{O}{3}] line ratio are not consistent with the low densities measured from the [\ion{Ne}{5}] lines in the IR.

Therefore, if the [\ion{Ne}{5}] emission comes from a similar distribution as the [\ion{O}{3}] as implied by the line profiles, then the [\ion{O}{3}] must arise in very dense clumps embedded within a much more diffuse, highly ionized plasma, from which originates the [\ion{Ne}{5}]. This conclusion is supported by the very early detection of permitted nitrogen emission \citep[e.g.,][]{IijimaEsenoglu03} as the ejecta entered into the nebular stage of evolution. Indeed, the inferred clump density is as high as was derived during the first 100 -- 300 days after eruption \citep[(5.8 -- 9.4)$\times 10^{6}$ cm$^{-3}$;][]{IijimaEsenoglu03}. This suggests that the clumps have experienced very little dissipation during the intervening 1300 -- 1500 days. Clumpiness in the ejecta of CNe seems to be relatively common as indicated by direct imaging of novae shells \citep[see][for examples]{OBrienBode08}, by photoionization modeling of the nebular emission \citep[e.g.,][]{Vanlandingham05,Helton10a}, and by examination of emission lines (e.g., V382 Vel; \S\ref{sec:V382Vel_Line_Struct}). In V1494 Aql the strong contrast in density between the diffuse and clumpy material is unusual. Photoionization models typically suggest a density contrast of a factor of $<10$, not $\sim 10^{4}$ as implied by the line ratios here.

The assumption that the temperature distribution is continuous across the boundary between the clumpy and diffuse material allows us to place constraints on $T_{e}$ using the line ratios. As demonstrated by the [\ion{O}{3}] ratio, $T_{e}$ must be greater than 7500 K. \citet{IijimaEsenoglu03} derived an electron temperature consistently around 11000 K over four observations during the first 100 -- 300 days of development (see below). The relatively high temperatures implied by the [\ion{O}{3}] lines suggest then that the temperature was ``frozen-in'' early in the evolutionary development. This is somewhat surprising considering the strength of the [\ion{O}{4}] 25.91 \micron\ line, which is typically an efficient IR coolant. However, the saddle-shaped [\ion{O}{4}] line profile differs strongly from that of the rounded O and Ne lines. If this is due to a different distribution of the [\ion{O}{4}] emitting material, then it might explain the high values of $T_{e}$ in spite of the presence of strong IR cooling lines. Indeed, Figure \ref{fig:FluxEvolve} indicates the the [\ion{O}{4}] line decays as $t^{-1.9}$, far too shallow to arise from a thin shell. It is possible that the distribution is in an equatorial ring as hypothesized by some authors to explain saddle-shaped emission profiles observed early in the outburst. It is interesting, however, that the other lines no longer show the same profile. This may imply that the [\ion{O}{4}] emission is coming from the remnants of the expanding ring structure.

\subsubsection{V1494 Aql - Adopted Parameters}
\label{sec:V1494_AV}

Estimates of the distance to V1494 Aql range from $\sim 1$ to 3.6 kpc \citep[Table \ref{tab:distances}; and see also][]{Helton10b}. For our abundance calculations, we used a distance of 1.6 kpc.

Several authors provide estimates of the mass of hydrogen in the ejecta. \citet{IijimaEsenoglu03} found that over a 150 day period, the electron temperature stayed nearly constant at T$_{\mathrm{e}} = 10,700 \pm 600$ K. The electron density, on the other hand, was found to decay $\propto t^{-0.8}$, from $9.4 \times 10^{6}$ cm$^{-3}$ at day 164 to $5.2 \times 10^{6}$ cm$^{-3}$ at day 289. Using those values, Iijima \& Eseno\v{g}lu estimated a hydrogen mass of $(4.7 \pm 1.0) \times 10^{-5}$ M$_{\odot}$. \citet{Eyres05} found N$_{\mathrm{e}} = (1.2 \pm 0.4) \times 10^{6}$ cm$^{-3}$ using the ratio of [\ion{O}{3}] $\lambda\lambda$4959,5007 \AA\ to [\ion{O}{3}] $\lambda$4363 \AA\ using a temperature of T$_{\mathrm{e}} = 10700$ K, which resulted in a hydrogen mass of m$_{\mathrm{H}} \sim 1.8 \times 10^{-5}$ M$_{\odot}$ for an assumed distance of 1.6 kpc. \citet{Kamath05} estimated the electron number density at day 510 to be N$_{\mathrm{e}} = (1.1 \pm 0.1) \times 10^{5}$ cm$^{-3}$ for T$_{\mathrm{e}} = 1.5 \times 10^{4}$ using the H$\beta$ line luminosity and assuming a spherical shell expanding at 2500 km s$^{-1}$ with a filling factor of 0.01. These parameters yielded m$_{\mathrm{H}} = 6 \times 10^{-6}$ M$_{\odot}$. We use the mean of these three mass estimates, $2.5 \times 10^{-5}$ M$_{\odot}$, for our abundance analysis.

The measured ejecta velocities varied as the system evolved. We take v$_{max}$ to be the highest velocity P-Cyg absorption component of 1850 km s$^{-1}$ and use this as the basis for the first ejecta model - a shell with the outer edge expanding at v$_{max}$ and with a depth 10\% of the radius, $\Delta r = 0.1 r_{out}$. This geometry results in n$_{e} \sim 1450$ cm$^{-3}$. We base the second model on the ellipsoidal geometry determined from 6 cm semi-resolved images by \citet{Eyres05}, who estimated the expansion velocities to be 980 km s$^{-1}$ and 2500 km s$^{-1}$ along the minor and major velocities respectively. Following their example, we chose v$_{exp}$ of the third axis to be the mean of the other two axes. Adopting a shell depth of 10\%, we find an electron density 1594 days post-outburst of n$_{e} \sim 650$ cm$^{-3}$. As discussed above, the optical emission lines indicate a high degree of density fragmentation in the ejecta. In spite of that, these two models are quite consistent with the densities expected from the IR [\ion{Ne}{5}] line ratios.

As mentioned above (\S\ref{sec:V1494Aql_Line_Struct}), the [\ion{O}{3}] 4363/(4959 + 5007) line ratio (Figure \ref{fig:V1494Aql_TemDen}) clearly indicates that the [\ion{O}{3}] emission arises in regions of very high density. Therefore, we conduct our abundance analysis on the clumps using densities of $10^{6}$ and $10^{7}$. In addition, the [\ion{O}{3}] line ratio also reveals that $T_{e}$ is not less than 7500 K. In fact, if we assume that the density of the clump material declined slightly during the $\sim1400$ days separating our earliest optical observations and those of \citet{IijimaEsenoglu03} and that it is unlikely that $T_{e}$ would increase significantly during that period, then the temperature is probably closer to 10000 K. Since we expect the temperature to be nearly uniform throughout both the clumpy and diffuse regions, we constrain $T_{e}$ to 10000 K for our abundance estimates in both. 

The adopted parameters are summarized in Table \ref{tab:parameters}.

\subsubsection{V1494 Aql - Abundances}
\label{sec:v1494_Ab}

Our abundance calculations yield relatively modest enhancements of Ne and Mg relative to solar in the diffuse component of the ejecta. Both values are relatively sensitive to the density assumed, with Ne ranging from 4.7 to 9.6 times solar and Mg from 3.4 -- 7.5 times solar. We estimate the minimum sulfur abundance to be 0.5 times solar by number with respect to hydrogen. As mentioned above (\S\ref{sec:V382_Ab}), a lower limit on the abundance that is below solar values does not provide useful constraints on the relative enrichment of the material in question. 

The observations of both [\ion{O}{3}] and [\ion{O}{4}] enabled us to calculate the abundance in both the diffuse and the dense components of the ejecta. We would expect that the two components would have similar abundances, and this is what is found. Since the optical data also include emission from hydrogen, we were able to estimate the abundances directly rather than relating the observed fluxes to an assumed ejecta mass as required for the IR line diagnostics. In this case, the dense component was revealed to have oxygen abundances of 3.1 -- 20.0 times solar. We can then compare this estimate to that for the diffuse component, which yields an abundance of 13.5 -- 24.8 times solar.  It is also important to recall that the line profile of [\ion{O}{4}] suggests an emitting region having an equatorial ring/polar cap morphology. If correct, this would imply that the [\ion{O}{4}] arises in regions with slightly higher densities than assumed for the other IR lines and would drive the derived abundances down somewhat, bringing them into closer alignment with those determined for the dense component. That these two independent estimates for the oxygen abundance are consistent is reassuring.

The results of our abundance analyses are presented in Table \ref{tab:v1494_ab}.  


\begin{deluxetable*}{lcccc}
\tabletypesize{\scriptsize}
\setlength{\tabcolsep}{0.025in}

\tablewidth{0pt}
\tablecaption{Abundances by Number Relative to Solar\tablenotemark{a} - V1494 Aql \label{tab:v1494_ab}}

\tablehead{
\colhead{Species}& \colhead{Wavelength}& \colhead{$T_{e}$ (K):}& \multicolumn{2}{c}{$10^{4}$}\\
\cline{4-5}\\[-1.5ex]
\colhead{}& \colhead{(\micron)}& \colhead{$n_{e}$ ($cm^{-3}$):}& \colhead{ \ 650 \ }& \colhead{ \ 1450 \ }
}

\startdata

[Ne V]& 14.32& & -3.58& -3.91\\

& 24.32& & -3.29& -3.60\\

[Ne VI]& 7.64& & -4.32& -4.67\\[0.5ex]

\hline\\[-1.5ex]

\textbf{Total Neon}& & & -3.09 (9.6)& -3.40 (4.7)\\[0.5ex]

\hline\\[-1.5ex]

[Mg V]& 5.61& & -3.60& -3.95\\

[Mg VII]& 5.50& & -4.34& -4.68\\[0.5ex]

\hline\\[-1.5ex]

\textbf{Total Magnesium}& & & -3.52 (7.5)& -3.87 (3.4)\\[0.5ex]

\hline\\[-1.5ex]

[O III]& 0.5007& & -2.81\tablenotemark{b} (3.1)& -2.01\tablenotemark{c} (20.0)\\
 
[O IV]& 25.91& & -1.91 (24.8)& -2.18 (13.5)\\

[S IV]& 10.51& & -4.83 (1.1)& -5.16 (0.5)\\[-2.5ex]

\enddata

\tablenotetext{a}{The number given is the log abundance by number relative to hydrogen. The number in parentheses is the abundance by number relative to hydrogen, relative to solar assuming the solar values of \citet{Asplund09}.}
\tablenotetext{b}{This was calculated assuming a clump density of $10^{6}$ cm$^{-3}$.}
\tablenotetext{c}{This was calculated assuming a clump density of $10^{7}$ cm$^{-3}$.}

\end{deluxetable*}


Few other estimates of the ejecta abundances of V1494 Aql exist in the literature. Estimates of the helium abundance by number relative to hydrogen by \citet{IijimaEsenoglu03} and \citet{Kamath05} ranged from 1.5-2.8 times solar. Additionally, \citet{Kamath05} estimated that the oxygen and sulfur abundances were both depleted with respect to hydrogen at $\sim 0.03$ times solar. These estimates are orders of magnitude smaller than the lower limits determined here. They also disagree with calculations of the oxygen abundance by \citet{Rohrbach09} who found an overabundance of oxygen ($\sim 17 - 35$ times solar) by modeling X-ray spectra. In light of the fact that Rohrbach et al.\ admit that their values are an overestimate, we consider their values to be in broad agreement with our IR derived O abundance.

\section{Discussion}
\label{sec:Discussion}

Based on the phenomenology of the outburst, one would expect that V1494 Aql was an ONe nova. It was a bright nova with a very fast light curve decline (t$_{2} < 7$ days), high expansion velocities ($> 2500$ km s$^{-1}$), and rapid transition to the coronal stage with emission from species having ionization potentials in excess of 400 eV -- properties that are all consistent with an ONe eruption. However, our derived abundances for O and Ne are not high enough to state with certainty that the outburst of V1494 Aql arose on an ONe WD, as these systems are expected to have Ne abundances $\gtsimeq 20$ times solar \citep{LivioTruran94,Schwarz07a}. 

\begin{figure*}
\epsscale{1.2}
\plotone{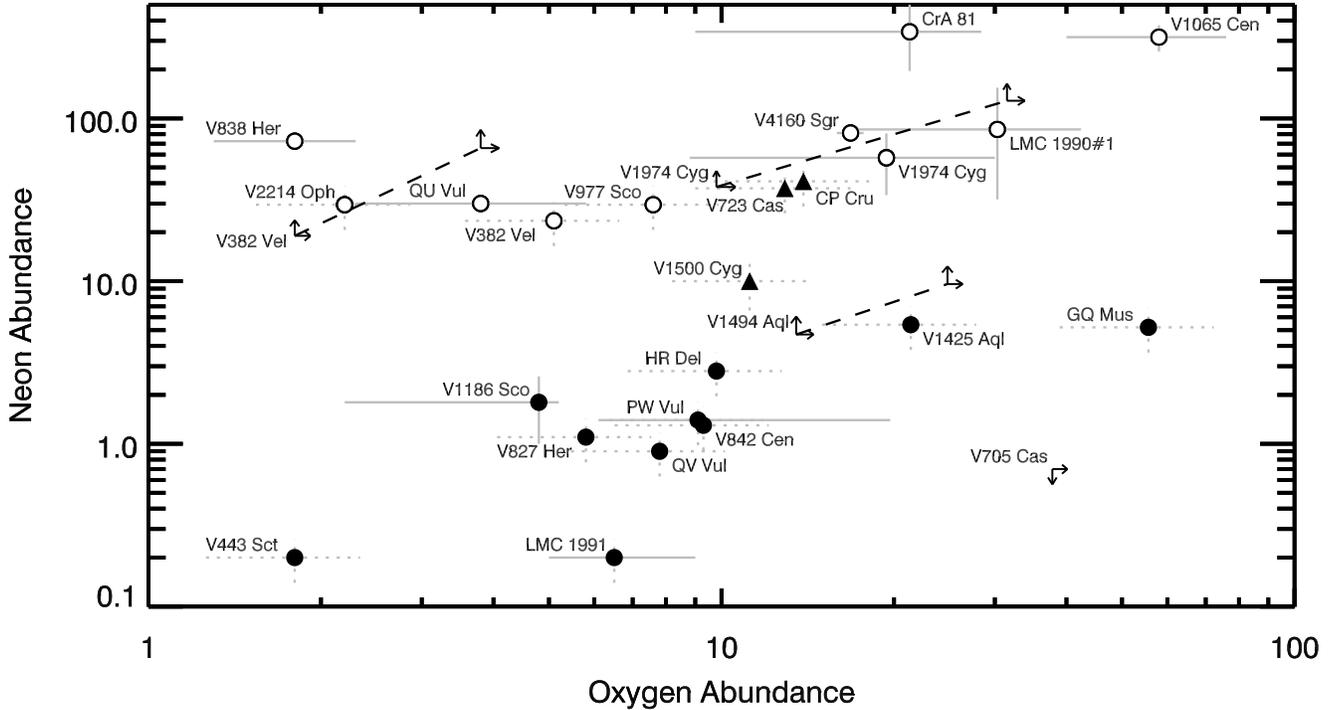}
\caption{Plotted above are oxygen vs.\ neon abundances taken from the literature for a selection of ONe and CO novae. The abundance values are reported by number relative to solar, where the data have all been renormalized to the solar values of \citet{Asplund09}. The open circles represent the ONe targets while the filled circles represent the CO novae. The filled triangles indicate the abundances of atypical CNe. See text for details. Solid error bars indicate the formal errors reported in the literature, when available. For those abundances with no reported errors, we assumed an error of $\pm 30$\% and indicated this with dashed error bars. The abundances for the CO nova V705 Cas include a lower limit to the O abundance and an upper limit to the Ne abundance. The lower limits determined in this work for V1974 Cyg, V382 Vel, and V1494 Aql are also indicated with the range of estimates derived using different environmental condistions linked by dashed lines. REFS --  QV Vul, V433 Sct, V827 Her, V842 Cen, V977 Sco, V2214 Oph - \citet{Andrea94}; V1065 Cen - \citet{Helton10a}; V723 Cas - \citet{Iijima06}; V1500 Cyg - \citet{Lance88}; CP Cru - \citet{Lyke03}; V1425 Aql - \citet{Lyke01}; GQ Mus - \citet{MorissetPequignot96}; V705 Cas - \citet{Salama99}; QU Vul - \citet{Schwarz02}; V838 Her, V4160 Sgr - \citet{Schwarz07a}; V1186 Sco - \citet{Schwarz07b}; PW Vul - \citet{Schwarz97}; LMC 1991 - \citet{Schwarz01}; V382 Vel - \citet{Shore03}; HR Del - \citet{Tylenda78}; V1974 Cyg - \citet{Vanlandingham05}; LMC 1990\#1 - \citet{Vanlandingham99}; CrA 81 - \citet{Vanlandingham97}. \label{fig:novaecompare}}
\end{figure*}

In Figure \ref{fig:novaecompare}, we compare the abundances of O and Ne reported in the literature for a selection of novae of both the ONe and CO types (open and closed circles respectively). Though we emphasize that the selection of targets for the plot is not exhaustive, it is nevertheless immediately apparent that there is a clear demarcation between the ONe- and CO-type systems. Filled triangles represent systems that have been reported in the literature as being peculiar in some way. V1500 Cyg was one of the fastest and brightest novae observed \citep{HachisuKato06}, yet its abundances seem to be intermediate between those of the ONe and CO novae classes \citep{Lance88}. Conversely, while the abundances of V723 Cas are considered to be typical of ONe novae, it had an exceptionally slow light curve development \citep[t$_{3} \sim 230$ days,][]{Iijima06}. In addition, it is highly atypical for an ONe nova since it has one of the longest known SSS stages among CNe, which is still ongoing after $> 15$ years \citep{Schwarz11}. \citet{Lyke03} classified CP Cru as a system intermediate between CO- and ONe-type novae based in part on its low Mg abundance relative to Ne.

The extremes of the abundance limits estimated for each of the sources examined in the present study are are shown linked with dashed lines. The range of abundances found for V382 Vel and V1974 Cyg are roughly consistent with what was determined through more detailed methods by other authors and both fall well within the range of abundances found for typical ONe novae. Further, the range of values determined in our analysis follow the trends observed across the more broad distribution of ONe sources. 

In contrast, the range of estimates for V1494 Aql clearly lie within the distribution of CO novae. In order for V1494 Aql to be considered an ONe nova, one would need to assume that the lower limits found for the neon abundance in the present study are an order of magnitude lower than the true abundances. This claim seems difficult to justify considering the lack of strong neon in the optical spectra and the broad consistency with other estimates found for V382 Vel and V1974 Cyg using the same methodology. Hence, we consider the abundances in V1494 Aql to be more consistent with an extreme CO nova, in spite of its ONe-like outburst properties. In this respect, V1494 Aql may be similar to very fast and bright novae such as LMC 1991 \citep{Schwarz01} or V1500 Cyg \citep{Lance88}. Both of these systems exhibited extreme outburst characteristics relative to their rather low neon abundances (less than solar in the case of LMC 1991 and $\sim 10$ times solar for V1500 Cyg).

Careful examination of Figure \ref{fig:novaecompare} also reveals a possible discrepancy between the results obtained in the present study for V1974 Cyg and those for the prototypical ONe nova QU Vul. The estimates for the neon abundances in both systems are high. In the case of V1974 Cyg, estimates range from 15 times solar \citep{Shore97} to $\sim 42$ times solar \citep{Vanlandingham05}. Our estimates here result in abundances of neon that are $> 35$ times solar. By comparison, QU Vul was found to have neon abundances ranging from $\sim 30$ times solar \citep{Schwarz02} to $> 65$ times solar \citep{Gehrz08a} (both scaled to the solar abundances of \citet{Asplund09}). Strong emission from [\ion{Ne}{2}] and [\ion{Ne}{3}] were observed in the IR at 7078 days after outburst in the ejecta of QU Vul. Yet our observations of V1974 Cyg revealed that the neon emission present at $\sim 4300$ days  after outburst had all but disappeared by day 5354. If the abundances were so similar, why would the neon emission in V1974 Cyg fade so much more rapidly than in QU Vul?

The differences cannot be due to the relative distances of the objects as the best estimates for both are $\sim 2$ -- 3 kpc. In addition, the final two \textit{Spitzer} observations of V1974 Cyg went significantly deeper than those of QU Vul -- up to 140 and 300 seconds for the SL and SH modules for V1974 Cyg versus 30 and 36 seconds for QU Vul -- implying that neither could the discrepancy be due to the depth of the observations.

We consider the best estimate for the ejecta mass of V1974 Cyg to be $2 \times 10^{-4}$ M$_{\odot}$ (\S\ref{sec:v1974_vals}), whereas the mass of ejecta in QU Vul is thought to be of order $3.6 \times 10^{-4}$ M$_{\odot}$ \citep[][and references therein]{Gehrz08a}. Again, this is not discrepant enough to account for the relative emission line strengths in the two systems. However, when one also considers the relative velocities of the two systems, a solution presents itself. The average ejecta velocity of V1974 Cyg was $\sim 2100$ km s$^{-1}$, nearly double that of QU Vul at $\sim 1200$ km s$^{-1}$. Hence, the dilution of the ejecta in V1974 Cyg would occur much more rapidly than in QU Vul. Since the emissivity of the emission lines is proportional to the square of the density, the difference in line strengths can be explained by the higher velocities in the ejecta of V1974 Cyg.

\section{Conclusions}
\label{sec:Conclusions}

We have presented IR and optical observations of three CNe in their late nebular stages ($> 4$ years post-outburst). We derived lower limits to abundances by number, relative to hydrogen using ejecta masses obtained during earlier stages of the outbursts. In the case of V1494 Aql, we used nearly contemporaneous optical spectra to obtain an independent estimate of the O abundance through direct comparison to H$\beta$. 

We placed lower limits on the abundances of oxygen and neon (9 and 35 times solar, respectively) in the ejecta of V1974 Cygni and found that the high abundances of these elements is consistent with earlier studies. Our derived abundances support the conclusion that the system contains an ONe WD. Although the emission lines of neon were weak relative to the [\ion{O}{4}] line at 25.91 \micron, the profiles suggested that the oxygen and neon emission may arise in different regions of the ejecta. Further, the stability of the oxygen line profile implied that the ejecta were still expanding ballistically and had not undergone substantial deceleration by the surrounding ambient medium.

The late time spectra of V382 Vel revealed emission lines from oxygen, neon, argon and sulfur, allowing us to estimate lower limits to the abundances of each of these elements. In particular, we found that the neon abundance was at least 18 times solar, with respect to hydrogen, oxygen was at least 1.7 times solar, and argon and sulfur were at least 0.1 times solar. The neon and oxygen abundances were consistent with results from previous studies. Our values represent the first abundance estimates for argon or sulfur in the ejecta of V382 Vel, but unfortunately they do not significantly constrain the abundances of these species. Examination of the emission line structure during this late optically thin stage demonstrated that the various emitting species had different spatial distributions in the ejecta and indicate the possible presence of ionization gradients. 

Our optical and \textit{Spitzer} IR observations of V1494 Aql provided the first detection of neon in this object. Based upon these data, we derived a minimum abundance of neon relative to hydrogen, relative to solar of about 5. This neon abundance is not high enough to confirm an ONe progenitor WD and, taken in concert with the high oxygen abundance ($\gtsimeq 14$ times solar), more strongly suggests a CO WD. The late time observation of neon in the ejecta of V1494 Aql reveals the importance of high sensitivity observations in the IR and near-UV regimes as well as the value of late time observations for characterizing the physical conditions and geometry of the nebular environment. Lower limits to the abundance of magnesium and sulfur were found to be 3.4 and 0.5 times solar, respectively.

\acknowledgements

LAH, CEW, and RDG were supported in part by NASA/JPL Spitzer grants 1289430, 1314757, 1267992, 1256406, and 1215746 to the University of Minnesota, The United States Air Force, as well as various National Science Foundation grants. SS acknowledges partial support from NSF, NASA, and \textit{Spitzer} grants to ASU.

{\it Facilities:} 
\facility{Spitzer (IRS)}, 
\facility{Bok (B\&C spectrograph)},


\end{document}